\documentclass[11pt]{article}

\usepackage{url}
\usepackage{amssymb,enumerate}
\usepackage{amsmath,amsfonts}
\usepackage{amsthm}
\usepackage{latexsym}
\usepackage{mathrsfs}
\usepackage{color}
\usepackage{microtype}
\usepackage{geometry}

\usepackage{algorithm}
\usepackage{algorithmicx}
\usepackage{algpseudocode}
\usepackage{enumitem}
\usepackage{empheq}
\usepackage{xcolor}
\usepackage{hyperref}
\usepackage{multirow}
\usepackage{appendix}
\usepackage[capitalise,nameinlink]{cleveref}
\usepackage{lmodern} 

\usepackage[caption=false,font=footnotesize]{subfig}
\usepackage{booktabs,tabularx}

\usepackage{amsmath,amsfonts,bm}

\def\eqref#1{equation~\ref{#1}}

\def\1{\bm{1}}

\def\eps{{\epsilon}}

\DeclareMathAlphabet{\mathsfit}{\encodingdefault}{\sfdefault}{m}{sl}
\SetMathAlphabet{\mathsfit}{bold}{\encodingdefault}{\sfdefault}{bx}{n}

\def\rmdist{{\mathrm{dist}}}

\def\cL{{\mathcal{L}}}

\newcommand{\st}{\mathrm{s.t.}}

\DeclareMathOperator*{\argmin}{arg\,min}

\hypersetup{backref=false,       
    pagebackref=true,               
    hyperindex=true,                
    colorlinks=true,                
    breaklinks=true,                
    urlcolor= black,                
    linkcolor= blue,                
    bookmarks=true,                 
    bookmarksopen=false,
    filecolor=black,
    citecolor=blue,
    linkbordercolor=blue
}

\theoremstyle{plain}

\theoremstyle{definition}

\usepackage[normalem]{ulem} 
\usepackage{xcolor}

\usepackage{graphicx, color}
\usepackage{algorithm, algpseudocode} 
\usepackage{mathrsfs} 

\usepackage{lipsum}

\title{Federated Learning–Assisted Optimization of Mobile Transmission with Digital Twins}
\author{Mohammad Heydari\thanks{Department of Electrical and Computer Engineering, McMaster University, CANADA (email: \url{heydam3@mcmaster.ca}; \url{todd@mcmaster.ca}; \url{dzhao@mcmaster.ca}).}
\and Terence D.\ Todd$^*$ 
\and Dongmei Zhao$^*$ 
\and George Karakostas\thanks{Department of Computing and Software, McMaster University, CANADA (email: \url{karakos@mcmaster.ca}).}}

\date{
\today
}

\begin{document}
\maketitle
	
\begin{abstract}
A Digital Twin (DT) may protect information that is considered private to its associated physical system. For a mobile device, this may include its mobility profile, recent location(s), and experienced channel conditions. Online schedulers, however, typically use this type of information to perform tasks such as shared bandwidth and channel time slot assignments. In this paper, we consider three transmission scheduling problems with energy constraints, where such information is needed, and yet must remain private: minimizing total transmission time when (i) fixed-power or (ii) fixed-rate time slotting with power control is used, and (iii) maximizing the amount of data uploaded in a fixed time period. Using a real-time federated optimization framework, we show how the scheduler can iteratively interact only with the DTs to produce global fractional solutions to these problems, without the latter revealing their private information. Then dependent rounding is used to round the fractional solution into a channel transmission schedule for the physical systems. 
Experiments show consistent makespan reductions with near-zero bandwidth/energy violations and millisecond-order end-to-end runtime for typical edge server hardware. To the best of our knowledge, this is the first framework that enables channel sharing across DTs using operations that do not expose private data.
\end{abstract}


\medskip

\section{Introduction}
Digital Twins (DTs) are software versions of physical systems that mirror device behavior on their behalf. In mobile communications, DTs can assist with functions such as scheduling and resource allocation, bandwidth sharing, base station association and power control. Prior work has studied many of these problems. DT modeling is used to assist adaptive aggregation for IIoT~\cite{Sun2021DTIIoT} and DTs are integrated with deep-RL optimization for industrial federated learning (FL)~\cite{Yang2023DTIndustrial}. Reference~\cite{Zhang2024DTFL_MEN} provisions multi-FL services with reinforcement-learned bandwidth control. Base station association and radio/edge resources are jointly optimized in a DT-enabled hierarchical FL framework in Reference~\cite{He2023DTFL_HCN}. The study in~\cite{Wu2025DTFL_NOMA} considers DT-assisted FL over non-orthogonal multiple access (NOMA) with game-theoretic coordination to reduce latency and energy. Reference~\cite{Lu2021LowLatencyDTFL} introduces a DT wireless network with blockchain-empowered FL for low-latency edge association.

DT-assisted scheduling arises in many applications, including federated-learning where devices periodically upload model updates~\cite{mcmahan2017communication}, industrial IoT workloads that offload sensing and computation to edge servers~\cite{Liu2022AGV_IIoT_Offloading}, and vehicles that transmit perception data and offloaded tasks to roadside infrastructure~\cite{materwala2022energy}. Another example arises in mobile transmission where the data consists of DT updates, i.e., physical systems periodically transmit DT state updates to satisfy periodicity requirements that keep DTs synchronized, while third-party applications interact only with the DTs rather than the devices themselves~\cite{Vaezi2022NetworkingPerspective}.

As part of its functionality, a DT may protect information that is considered private to its associated physical system~\cite{Son2022PrivacyPreservingDT}. For a mobile device, this may include information such as the device's mobility profile, recent locations, and experienced channel conditions. Such channel parameters (e.g., channel gains/CSI or related propagation features) can implicitly encode location-dependent characteristics and may therefore enable location inference or tracking by an adversary. For example, prior work shows that an adversary can localize user equipment by exploiting location-relevant wireless channel parameters/CSI~\cite{LiMitraTSP2024FPI}, and that access to received-signal measurements (RSSI) can be sufficient to infer users' common locations over time~\cite{Cunha2025RSSIPrivacy}. Clearly this type of information is useful in resource allocation when it is used by a central coordinator to create bandwidth allocations, time-slot assignments, and other outputs that are needed for efficient network operation. But the DTs may be reluctant to directly share this type of raw information, in order, for example, to protect participants against membership inference and gradient-recovery attacks~\cite{Zhu2019DLG} in Machine Learning applications. This is the main concern addressed in this paper.

In this work, we address the open question of how to exploit DTs in joint physical system transmission scheduling and resource allocation without revealing raw information to the central coordinator. To address this problem, a framework is proposed that keeps sensitive data local, and DTs exchange only lightweight summaries that do not directly reveal private data. This is done via a decentralized, federated optimization scheme in which, in each iteration, each DT solves a local subproblem using private data while a central coordinator interacts with these clients, in order to finally converge to the optimal solution of the global optimization problem. Here, privacy is respected by design under standard federated optimization/learning assumptions~\cite{Kairouz2021FL} (i.e., raw client/DT data remain local), and we do not claim additional formal guarantees such as differential privacy or secure aggregation.

We consider three variants of the scheduling problem with energy constraints that differ by whether transmit power control is used. These are referred to as problems P1, P2 and P3 as follows: (P1) fixed-power transmissions with DT channel knowledge or predictions, aiming to minimize the transmission period; (P2) fixed-power transmissions within a given time block, so that the (common) fraction of data transmitted is maximized (thus minimizing the number of blocks needed to complete all data transmission); and (P3) fixed-rate time slotting with power control transmissions, with the same objective as P1. After fractional solutions to relaxations of these problems are computed using federated optimization as described above, integral schedules are obtained via dependent rounding.

\subsection{Our contributions}

The main contributions of the paper are summarized as follows.

\begin{itemize}[leftmargin=1em]

\item \textbf{Problem Formulations.} We formalize the three variants P1-P3 of the scheduling problem with energy constraints when fixed-power transmissions are used (P1, P2) or fixed-rate, single-user slotting with power control is utilized (P3).

\item \textbf{Federated learning (FL) based optimization.} The scheduling problems are mapped into a constrained federated workflow where each digital twin keeps raw data local and shares only lightweight summaries with a central coordinator. An FL-based optimization method handles per-user (local) and system-wide (global) constraints in a decentralized solving of convex relaxations, producing high-quality fractional decisions. This approach has two main advantages:
\begin{enumerate}
\item In prior DT-assisted scheduling approaches (e.g., \cite{Li2024DTResourceAlloc,Huang2024UDTResourceMgmt}) centralized access to client/DT private data was assumed in order to make scheduling decisions. To the best of our knowledge, our framework is the first one that supports channel scheduling without directly exposing raw client/DT data. 
\item Since the communication required between the DTs and the central coordinator (which may itself be one of the DTs) is done via the high speed wired backbone network, the communication overhead between DTs is small when compared to the wireless communication between DTs and their corresponding Physical Systems (PS). The overall latency is heavily dominated by the local algorithm execution at the DTs and coordinator, rather than by communication. This allows us to factor this overhead out of our analysis.
\end{enumerate}

\item \textbf{Solution framework.} Two schedulers are introduced: the Fixed-Power Scheduler (FPS) for the fixed-power variants and the Power-Control Scheduler (PCS) for the power-control variant. Each algorithm uses a pipeline that solves convex relaxations, converting fractional outputs into integral schedules via simple or smart dependent rounding algorithms, re-optimizing bandwidth for fixed-power cases when needed, and applying minimal feasibility bandwidth boosting for FPS and energy for PCS so that all constraints are met. These algorithms make the resource versus latency trade-off explicit and produce good results.

\item \textbf{Evaluation.} Simulations on wireless traces with FPS and PCS show consistent makespan reductions with negligible budget violations and end-to-end runtime on the order of milliseconds, suitable for edge deployment.

\end{itemize}

\subsection{Related work}

Existing networking work that is closest to ours falls into two streams.

\paragraph{Digital twins for edge scheduling and offloading.} A first stream leverages DTs to guide networked resource management at the edge. Zhang \emph{et al.} design a DT-driven framework for collaborative MEC task offloading, where virtual replicas steer low-latency decisions under dynamics \cite{Zhang2023JSAC_DT_Offloading}. Hao \emph{et al.} target URLLC and show that DTs can improve robustness under model mismatch while jointly minimizing delay and energy \cite{Hao2023JSAC_DT_URLLC}. Mobility and service-centric variants include DT-assisted offloading in UAV-aided edge networks \cite{Li2022TVT_DT_UAV_Offload} and DT-guided reinforcement learning for microservice offloading and bandwidth allocation across collaborative edges \cite{Chen2024DT_Microservice}. Broader surveys review DT-assisted task offloading in safety-critical systems and outline open challenges in modeling fidelity and closed-loop control \cite{Carmo2024Survey_DTOffload}. These efforts demonstrate that DTs can forecast state, evaluate candidate policies, and close the loop for resource management. Complementary to DT-assisted edge scheduling and offloading, scheduling and resource allocation for federated learning in vehicular networks have also been studied in~\cite{HeydariVTC2025} Our paper complements them by addressing uplink scheduling with privacy-preserving coordination between DTs and a central scheduler.

\paragraph{Federated optimization/learning for constrained optimization.} A second stream develops methods that solve optimization problems with both global and client-local constraints while keeping raw data on devices. He, Peng, and Sun propose a federated constrained-optimization framework using an augmented Lagrangian with inexact ADMM updates and provide complexity guarantees \cite{He2024FederatedConstraints}. Related distributed optimization foundations include projected-subgradient and primal–dual schemes for local or conic constraints \cite{Nedic2010TAC,Lin2016Automatica,Wang2017TAC,Aybat2016NIPS,Aybat2019SIOPT}, as well as primal–dual projected approaches that handle simultaneous global and local constraints \cite{Zhu2011TAC,Yuan2011TSMCB}, with surveys in \cite{Yang2019ARC}. There is also application-focused constrained FL on fairness and imbalance \cite{Shen2021ICLR,Chu2021FedFair,Du2021SDM,Galvez2021NeurIPSW}, though these typically target specialized objectives. Our work builds on the general FL-with-constraints paradigm to coordinate DTs and a central scheduler where sensitive coefficients remain local to the DTs while the system jointly optimizes uplink schedules and resource allocation.

\section{System Model}
\label{sec:systemmodel}

\begin{figure}[t]
  \centering
  \includegraphics[width=88mm]{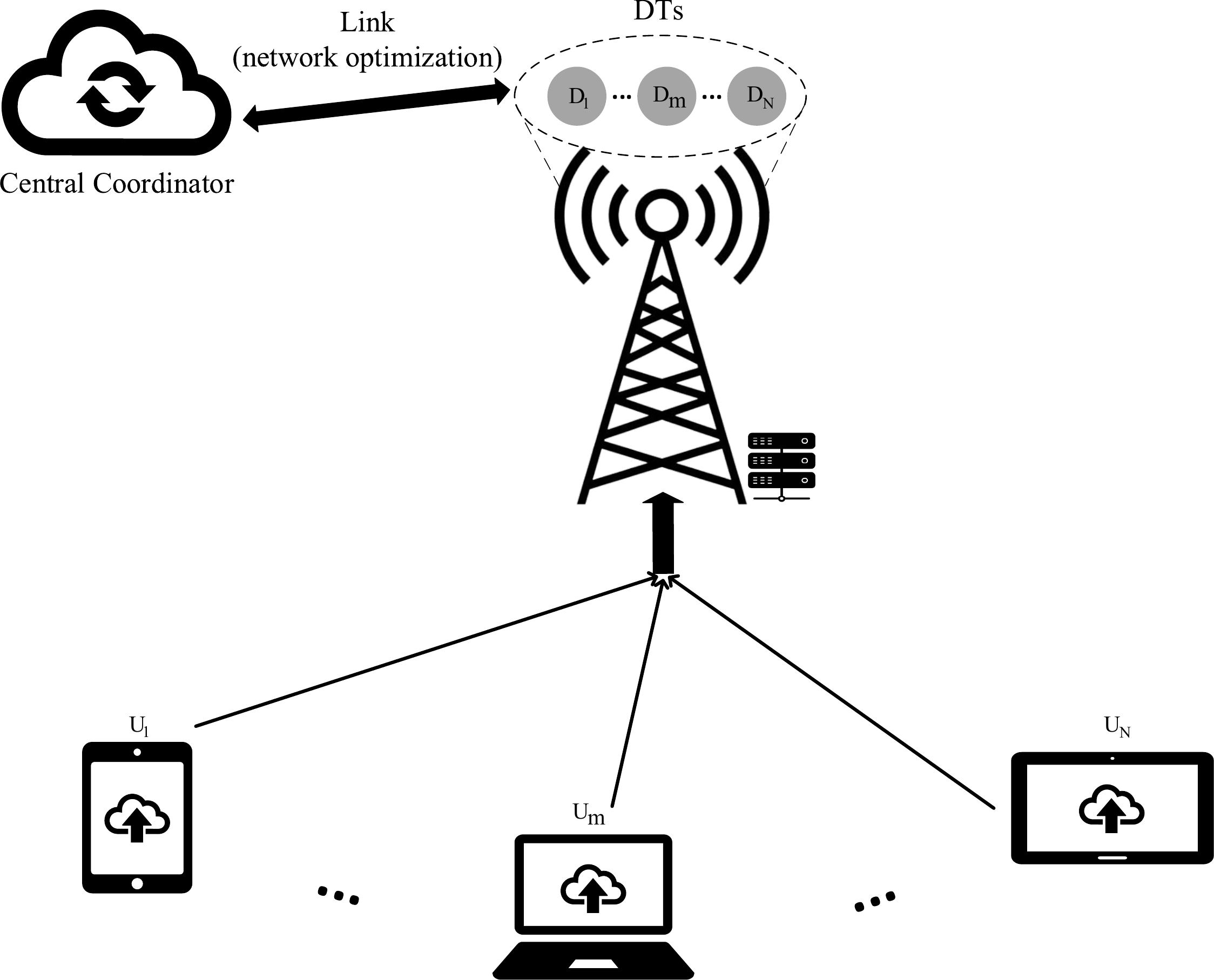}
  \caption{System model.}
  \label{fig:1}
\end{figure}

We consider a wireless communication network that facilitates the uplink transmissions of a set of mobile users. The scheduling that occurs is recurrent, and a central coordinator dynamically controls the use of a shared wireless channel so that users can upload their data. The communication infrastructure is shown in Fig.~\ref{fig:1}. It consists of a single base station (BS), a set of users and their corresponding DTs that are hosted at the edge of the network. The DTs act on behalf of their associated users and collaborate with the central coordinator in order to plan the uplink user transmissions.  In order to do so, each DT holds information that can enable the coordinator to make resource assignment decisions. This information includes recent uplink channel conditions, which are considered private to each user. This information may be updated between a given user and its DT as part of the usual digital twin synchronization process \cite{Chen2024}. Alternately, this information may be communicated directly by the base station to each DT. Note that this information is not directly shared with the central coordinator, but instead, it shares summary information as part of the federated optimization process discussed later.  We adopt a time-slotted wireless communication channel where transmission occurs within discrete time slots, each of duration $\tau$. At every time slot, the total available uplink BS bandwidth is $B$ Hz.

Let $\mathcal{N}=\{1,2,\ldots,N\}$ denote the set of users and $\mathcal{T}=\{1,2,\ldots,T\}$ the set of slots in a scheduling period. When needed, we also use fixed-length blocks of $t_s$ slots and write $\mathcal{T}_s=\{1,2,\ldots,t_s\}$ for a single block. The parameter $t_s$ represents the maximum common horizon over which all users can obtain channel knowledge or predictions. As a general guideline, the choice of $t_s$ should satisfy $t_s \tau < T_c$, where $T_c$ is the channel coherence time.

Our objective is to schedule the uplink transmission of tasks from all users via their DTs, which collaborate with the central coordinator, so as to minimize the \emph{makespan}, i.e., the time by which every user has completed their upload.
We consider three variants of the scheduling problem, distinguished by whether power control is enabled. These are referred to as Problems P1, P2 and P3. When power control is not used, i.e., each user transmits with a fixed power in Problems P1 and P2 (which differ in the available horizon of channel knowledge). When power control is enabled, we define Problem P3.

For each user $m \in \mathcal{N}$, the following quantities are defined:
\begin{itemize}
	\item $U_m$ is the edge device of user $m$ that communicates with the BS.
  	\item $D_m$ is the DT associated with user $m$, hosted at the edge server.
  \item $C_{m}$ is the size (in bits) of the task that user $m$ must upload to the BS.
  \item $\bar{P}_{m}$ is \emph{fixed} transmit power of user $m$ during one time slot (used in Problems~P1 and~P2).
  \item \textbf{$R$} is the target uplink rate (bit/s) prescribed by the scheduler when power-control is enabled (Problem~P3).
  \item \textbf{$P_{m,t}$} is \emph{slot-adaptive} power chosen by user $m$ in slot $t$ under power control in Problem P3, where the total bandwidth $B$ is assigned to a single user. To guarantee the fixed rate $R$, Shannon’s formula yields:
        \begin{equation}
          P_{m,t} = \frac{N_0 B}{g_{m,t}} \left(2^{\frac{R}{B}} - 1\right)
,\label{eq:pc}
        \end{equation}
        where $g_{m,t}$ is the link gain, $N_0$ is the noise power spectrum density at the BS receiver input, and $B$ is the available bandwidth.
  
  \item $E_{m}$ is the energy budget allocated to user $m$. The total energy consumption of user $m$ over all $T$ time slots must not exceed this budget.
\end{itemize} 

In each period, the DTs and the central coordinator collaborate to compute a feasible schedule for the upcoming transmissions. The resulting plan is disseminated to the users and executed at the start of the next period (or the next block of $t_s$ slots in P2). The optimizer runs with a one period look ahead: while the current plan is being executed, the DTs and the coordinator prepare the schedule for the following period. The system always remains under a valid schedule, since the last feasible plan stays in effect until a new one starts.
We assume that either the number $N$ of users that need scheduling simultaneously is small or the users can be partitioned into independent groups that can be separately scheduled.

\section{Problem Formulation}
\label{sec:formulation}

We study the problem of scheduling the periodic transmission of $C_m$ bits for each user $m$ over a period of $T$ time slots, and the determination of its uplink data rates for each time slot, so that period $T$ is minimized, without exceeding the available channel bandwidth $B$ at every time slot or the energy budget of each user.
   
We will consider three different variations of this problem. If there is no power control for the user transmissions, i.e., there is a fixed transmission power $\bar{P}_{m}$ for each user $m$, then we consider the following two variations, that differ in their assumption of channel knowledge and/or prediction by a user:
\begin{description}
\item[P1] We assume that each user has knowledge or prediction capabilities of channel parameters, such as gain factors, for a {\em large-enough time horizon} so that a (minimized) time period $T$ for the whole transmission is always possible. In this case, we minimize $T$.
\item[P2] Without the previous assumption, let $t_s$ be the maximum possible common time horizon during which all users can have knowledge or prediction capabilities of channel parameters. Typically, we assume that $t_s$ is significantly smaller than the time needed to finish the whole transmission (otherwise this problem reduces to problem P1). In this version of the problem, we break the transmission into blocks of $t_s$ time slots, and we minimize the number of blocks used, by maximizing the fraction of data transmitted by the users during a single block. 
\end{description}
The third version of the problem assumes the existence of power control for the user transmission. We denote this version by {\bf P3}.

For all three versions, let $x_{m,t} \in \{0,1\}$ be the decision variable that determines whether user $m$ uploads during time slot $t$ ($x_{m,t} = 1$) or not ($x_{m,t} = 0$). 

{\bf Energy budget constraints:} For P1 and P2, let $n_{m} :=\lfloor E_{m}/(\bar{P}_{m}\tau)\rfloor$ be the maximum number of time slots that user $m$ can utilize for data transmission due to its energy budget. Then the following constraints must hold for P1:
\begin{equation}  \label{eq:CPUO}
\sum\limits_{t = 1}^{T} x_{m,t} \leq n_m,\ \ \forall m\in \mathcal{N}. 
\end{equation}
For P2, if $\gamma$ is the fraction of data demand transmitted during a block $t_s$ of time slots, then it is natural to allocate only a fraction $\gamma$ of the energy budget $n_m$ to this transmission, for every user $m$. Then we have the following constraints:
\begin{equation}  \label{eq:CPUO2}
\sum\limits_{t = 1}^{t_s} x_{m,t} \leq \gamma n_m,\ \ \forall m\in \mathcal{N}. 
\end{equation}

In P3 the energy budget is enforced directly through adaptive power:
\begin{equation}  \label{eq:energy_pc}
\sum\limits_{t = 1}^{T} P_{m,t}x_{m,t}\leq \frac{E_m}{\tau},\ \ \forall m\in \mathcal{N}.
\end{equation}

{\bf Bandwidth constraints:} In P1 and P2, let $w_{m,t}$ be the decision variable that determines the {\em fraction} of total bandwidth allocated to user $m$ in time slot $t$. Then the following constraints must hold:
\begin{equation}   \label{eq:bandwidth_constraint}
 \sum\limits_{m= 1}^{N} w_{m,t} \leq 1,\ \ \forall t\in \mathcal{T}.
\end{equation}
To ensure that no bandwidth is assigned when user $m$ does not transmit during time slot $t$, we add the following constraints:
\begin{equation}
\label{eq:coupling_constraint}
w_{m,t} \le x_{m,t}, \ \
 \forall m\in \mathcal{N}, t\in \mathcal{T}.
\end{equation}

In P3, there are no bandwidth constraints, but since each slot is allocated to a single user, we must have:
\begin{equation}  \label{eq:single_user}
\sum\limits_{m = 1}^{N} x_{m,t} \leq 1,\ \ \forall t\in \mathcal{T}.
\end{equation}

{\bf Demand constraints:} The achievable uplink rate for user $m$ in slot $t$ is
\begin{equation}
\label{eq:rate_model}
r_{m,t}=\begin{cases}
B w_{m,t}\log_{2}\!\Bigl(1+\frac{\bar{P}_{m}g_{m,t}}
{N_{0}B w_{m,t}}\Bigr), & \text{for P1, P2}\\
R, & \text{for P3.}
\end{cases}
\end{equation}

During period $T$, each user $m$ must transmit $C_m$ bits. For P1, this implies the following demand constraints:
\begin{equation}  \label{eq:rateP1}
\sum_{t=1}^{T} r_{m,t} \geq \frac{C_m}{\tau},\ \ \forall m\in \mathcal{N}.
\end{equation}
and for P2 the following constraints:
\begin{equation}  \label{eq:rateP2}
\sum_{t=1}^{t_s} r_{m,t} \geq \gamma \frac{C_m}{\tau},\ \ \forall m\in \mathcal{N}.
\end{equation}
For P3, let $\lceil C_{m}/(R\tau)\rceil$ be the number of time slots needed by user $m$. Then the demand constraints become:
\begin{equation}  \label{eq:demand-p3} 
\sum_{t=1}^{T} x_{m,t} \ge \lceil C_{m}/(R\tau)\rceil,\ \ \forall m\in\mathcal{N}.
\end{equation}
  
In order to solve P1, we will perform a binary search for $T$ within a sufficiently large range. For each $T$ value, we will look for a feasible solution $\mathbf{x}\in \{0,1\}^{N\times T}, \mathbf{w}\in \mathbb{R}_+^{N\times T}$ that satisfies constraints \cref{eq:CPUO}, \cref{eq:bandwidth_constraint}, \cref{eq:coupling_constraint}, and \cref{eq:rateP1}. Depending on whether there is a feasible solution or not, we try a smaller or larger value for $T$, respectively. 

We formulate the non-linear (but convex) mixed-integer mathematical program we solve at each binary search step for P1. In order to align its formulation with the formulation of P2, we introduce here also the fraction $\gamma$ of user data transmitted in $T$ time slots, and, instead of feasibility, we check whether the maximum $\gamma$ is at least $1$ or not. Abusing the notation a bit, we will use the term ``infeasible" for the latter case, i.e., $\gamma<1$. Hence we solve the following:
\begin{align}
\max_{x, w, \gamma} & \ \gamma \text{   s.t.}  \label{Eq:1maxy} \tag{P1}  \\
&\sum\limits_{m = 1}^{N} w_{m,t} \leq 1,\ \ \forall t\in \mathcal{T}   \label{Eq:1C1} \\
&\sum\limits_{t = 1}^{T} B{\log _2}( 1 + \frac{{\bar{P}_{m}}{g_{m,t}}}{N_0Bw_{m,t}} )w_{m,t} \geq \gamma\frac{C_m}{\tau},\ \ \forall m\in \mathcal{N} \label{Eq:1C2} \\
&\sum\limits_{t = 1}^{T}  x_{m,t} \le n_m,\ \ \forall m\in \mathcal{N} , \label{Eq:1C3} \\
&w_{m,t} \leq x_{m,t},\ \ \forall m\in \mathcal{N}, t\in \mathcal{T}  \label{Eq:1C4} \\
&x_{m,t} \in \{0,1\},\ \ \forall m\in \mathcal{N}, t\in \mathcal{T}   \label{Eq:1C5} \\
&w_{m,t} \geq 0,\ \ \forall m\in \mathcal{N}, t\in \mathcal{T} \notag \\
&\gamma \geq 0.\ \  \notag
\end{align}
Our objective is to compute $T^*$, the minimum ${T}$ for which \ref{Eq:1maxy} has a feasible ($\gamma\geq 1$) solution.

Solving problem \ref{Eq:1maxy} is NP-complete by a simple reduction from {\sc Partition}. The reduction works by setting $B$ equal to exactly half the total value of the $n$ items in {\sc Partition}, $N:=n$, $T:=2$, $C_m$ equal to the item value (adjusted to account for the $\log_2$ factor in \cref{Eq:1C2}), and $n_m=1$ for every $m$. Then there is a solution to {\sc Partition} iff the items can be partitioned into the two available time slots each with exactly half the total value. 

For problem P2, there is no need for binary search since the time period $t_s$ is fixed. The formulation of P2 becomes
\begin{align}
\max_{x, w, \gamma} & \ \gamma \text{   s.t.}  \label{Eq:2maxy} \tag{P2}  \\
&\sum\limits_{m = 1}^{N} w_{m,t} \leq 1,\ \ \forall t\in \mathcal{T}_s  \label{Eq:2C1} \\
&\sum\limits_{t = 1}^{t_s} B{\log _2}( 1 + \frac{{\bar{P}_{m}}{g_{m,t}}}{N_0Bw_{m,t}} )w_{m,t} \geq \gamma\frac{C_m}{\tau},\ \ \forall m\in \mathcal{N} \label{Eq:2C2} \\
&\sum\limits_{t = 1}^{t_s}  x_{m,t} \le \gamma n_m,\ \ \forall m\in \mathcal{N} , \label{Eq:2C3} \\
&w_{m,t} \leq x_{m,t},\ \ \forall m\in \mathcal{N}, t\in \mathcal{T}_s  \label{Eq:2C4} \\
&x_{m,t} \in \{0,1\},\ \ \forall m\in \mathcal{N}, t\in \mathcal{T}_s   \label{Eq:2C5} \\
&w_{m,t} \geq 0,\ \ \forall m\in \mathcal{N}, t\in \mathcal{T}_s \notag \\
&\gamma \geq 0.\ \  \notag
\end{align} 

For P3, we again perform a binary search to find the minimum value $T^*$ for $T$ for which constraints \cref{eq:energy_pc}, \cref{eq:single_user}, and \cref{eq:demand-p3} are satisfied, i.e., the following is feasible:
\begin{align}
\min_{x} & \ 0 \text{   s.t.}  \label{Eq:3min0} \tag{P3}  \\
&\sum\limits_{m = 1}^{N} x_{m,t} \leq 1,\ \ \forall t\in \mathcal{T}   \label{Eq:3C1} \\
&\sum\limits_{t = 1}^{T} P_{m,t}x_{m,t}\leq \frac{E_m}{\tau},\ \ \forall m\in \mathcal{N} \label{Eq:3C2} \\
&\sum\limits_{t = 1}^{T}  x_{m,t} \geq \lceil C_{m}/(R\tau)\rceil,\ \ \forall m\in \mathcal{N} , \label{Eq:3C3} \\
&x_{m,t} \in \{0,1\},\ \ \forall m\in \mathcal{N}, t\in \mathcal{T}   \label{Eq:3C4} 
\end{align}
\ref{Eq:3min0} is also NP-complete by a reduction of the well-known  {\sc Partition} problem to it. In the latter problem, we are given $n$ values $a_1, a_2,\ldots, a_n$ and we ask whether there is a partition into two sets, each of cardinality $n/2$, so that the sum of values in each is exactly $\sum_{i=1}^na_i/2$. The reduction sets $m:=2$ (i.e., two users), $T:=n$, $\frac{E_m}{\tau}:=\sum_{i=1}^na_i/2$, $P_{1,t}=P_{2,t}:=a_t$, and $\lceil C_{m}/(R\tau)\rceil:=n/2$. Then it is easy to see that \ref{Eq:3min0} has a solution iff the {\sc Partition} problem has one.

\section{Solution Algorithms}
\label{sec:solution}

\begin{figure*}[ht!]
    \centering 
    \includegraphics[width=\textwidth]{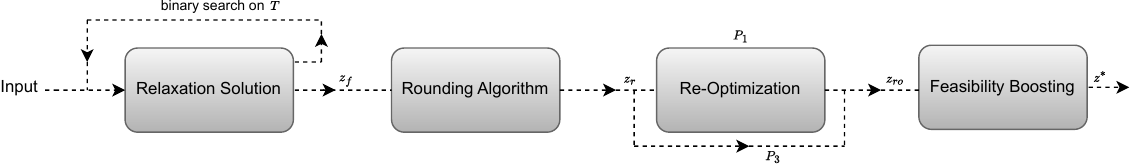}
   	\caption{General solution scheme.}
   	\label{solution_pipeline}
\end{figure*}

Since the solution of problem P2 is very similar to solving problem P1, in what follows we are going to concentrate on solving P1 and P3.

Our general solution scheme for both problems is outlined in Figure~\ref{solution_pipeline}:

{\bf Binary search:} Since all problems are NP-complete, we do not expect to have an efficient algorithm that solves them. Therefore, we are going to solve their {\em relaxations}, obtained by relaxing constraints $x_{m,t}\in\{0,1\}$ into $x_{m,t}\geq 0$, and then rounding the obtained fractional solution. Note that for \ref{Eq:1maxy} and \ref{Eq:2maxy}, if the relaxation is feasible we can always have a feasible solution with $x_{m,t}\leq 1\ \forall m,t$ by reducing any $x_{m,t}>1$ to $x_{m,t}=1$ without violating any constraints, since~\cref{Eq:1C1} for \ref{Eq:1maxy}, and \cref{Eq:2C1} for \ref{Eq:2maxy} guarantees that $w_{m,t}\leq 1\ \forall m,t$, so there is no need for $x_{m,t}>1$.

We perform a binary search on the value of $T$, starting with $T:=1$, as follows: We double the current value of $T$ until the relaxation becomes feasible; then we perform a usual binary search between the previous and the current $T$ values, in order to obtain the smallest value $T_f$ that allows relaxation feasibility. Let $z_f=(x_f, w_f)$ or $z_f=x_f$ be the feasible fractional solution of \ref{Eq:1maxy} or \ref{Eq:3min0} corresponding to $T_f$, respectively. 

{\bf Rounding:} The fractional solution $z_f$ is rounded to a new solution $z_r$, so that a transmission schedule $x_r$ is produced ($x_r(m,t)\in \{0,1\}$). The {\em dependent rounding} method we use produces a schedule $\mathbf{x}$ that respects the integral constraints, i.e., the energy bound \cref{Eq:1C3} of \ref{Eq:1maxy} and demand bound \cref{Eq:3C3} of \ref{Eq:3min0}, but it may violate constraints \cref{Eq:1C2} or constraints \cref{Eq:3C2}.

{\bf Re-optimization for P1:} Keeping fixed the rounded schedule $\mathbf{x}_r$ as calculated in the previous step, we re-solve \eqref{Eq:1maxy}, to obtain possibly better bandwidth allocations $\mathbf{w}_{ro}$ for P1.

{\bf Feasibility boosting:} If there are violated constraints \cref{Eq:1C2} of \ref{Eq:1maxy}, then we increase the available bandwidth $B$ until they are all satisfied. Similarly, if there are violated constraints \cref{Eq:3C2} of \ref{Eq:3min0}, then we increase the available energy $E_m$ until they are all satisfied. An alternative solution would be to avoid scheduling the whole demand of users with violated constraints during the current period $T_f$, but postpone the transmission of the remaining demand in subsequent periods. We do not adopt this solution, since our solution ensures the {\em synchronicity} of all DT transmissions, which can be a requirement for the application using the transmitting DTs.   

While the solution method summarized above is easily implementable if complete information is available to all DTs, and no time restrictions are imposed, the requirements of {\em data privacy} and {\em periodic transmission} in this work preclude such a straight-forward approach. Instead, we propose solutions that address both these requirements.

\subsection{Data privacy via Federated Learning}  
\label{sub:privacy}

We assume that certain pieces of data (coefficients) of relaxations \ref{Eq:1maxy} or \ref{Eq:3min0} are {\em private}, i.e., they are known only to their users and the corresponding DTs. We assume that data $Z^{P1}_m := \{C_m, n_m, \bar P_m, [g_{m,t}]_{t=1}^T\}$ for P1, or data $Z^{P3}_m := \{C_m, E_m, [P_{m,t}]_{t=1}^{T}\}$ for P3, are private to user $m$, since the latter does not want to disclose any information about their location, distance from the BS, channel conditions, battery status, etc. Note that, although channel data for all users are available at the BS, access to these data is only permitted to the DTs of their owners. 

Equivalently, we assume that constraints \cref{Eq:1C2}-\cref{Eq:1C3} in \ref{Eq:1maxy}, and \cref{Eq:3C2}-\cref{Eq:3C3} in \ref{Eq:3min0}, are {\em local} for each user $m$. In order to solve \ref{Eq:1maxy} or \ref{Eq:3min0} while keeping the coefficients of these constraints private, we employ the recent Federated Learning (FL) algorithm of~\cite{He2024FederatedConstraints}, where the clients are the user DTs, and the global model to be learned is a solution to \ref{Eq:1maxy} or \ref{Eq:3min0}, residing in the central coordinator. 

\cite{He2024FederatedConstraints} solve general convex optimization problems with global and local constraints of the following form:
\begin{align}   \label{eq:FL_Framework}
\min_{\phi} \quad & \sum_{m=1}^{N} f_m(\phi; Z_m) + h(\phi), \\[2mm]
\text{subject to} \quad & c_0(\phi;Z_0) \le 0, \\[2mm]
& c_m(\phi;Z_m) \le 0,\quad \forall m\in \mathcal{N}. 
\end{align}
where
\begin{itemize}
  \item $\phi \in \mathbb{R}^d$ is the decision variables vector,
  \item $f_m(\phi;Z_m)$ is the objective function for client $m$ with data $Z_m$,
  \item $h(\phi)$ is a closed convex regularization function,
  \item $c_0(\phi; Z_0)$ are global constraints with central coordinator data $Z_0$, and
  \item $c_m(\phi; Z_m)$ are local constraints (unique to each client).
\end{itemize}
The clients enter into rounds of interaction with the central coordinator, where each of the clients (users) and the central coordinator maintains a set of Lagrangian multipliers for all the constraints (corresponding to the `model' notion of FL). They enter into rounds of interaction, where in each round the clients send their current estimates of the Lagrangian multipliers (their `models') to the central coordinator, the latter computes a new solution $\phi$ and new estimates for its own Lagrangian multipliers (its `model'), and sends the new $\phi$ to the clients. The interaction terminates when the maximum divergence between successive solutions falls below a threshold parameterized by parameter $\eps_1$ {\em and} the maximum divergence between successive model estimates falls below a threshold parameterized by parameter $\eps_2$. 

To apply the algorithm of \cite{He2024FederatedConstraints}, we set 
\begin{itemize}
\item $\phi:=(\mathbf{x}, \mathbf{w}, \gamma)$ for \ref{Eq:1maxy} or $\phi:=\mathbf{x}$ for \ref{Eq:3min0}.
\item $f_m(\phi;Z_m):=0, \forall m$ for both \ref{Eq:1maxy}, \ref{Eq:3min0} and $Z_m$ defined as above. 
\item $h(\phi):=-\gamma$ for \ref{Eq:1maxy} and $h(\phi):=0$ for \ref{Eq:3min0}. 
\item Global constraints $c_0(\phi; Z_0)$ (where $Z_0=\emptyset$ in our case) are constraints \cref{Eq:1C1} and \cref{Eq:1C4} for \ref{Eq:1maxy}, and constraints \cref{Eq:3C1} for \ref{Eq:3min0}.
\item As mentioned above, local constraints $c_m(\phi; Z_m)$ are constraints \cref{Eq:1C2}-\cref{Eq:1C3} for \ref{Eq:1maxy}, and \cref{Eq:3C2}-\cref{Eq:3C3} for \ref{Eq:3min0}.
\end{itemize}

In what follows, we describe the Federated Learning (FL) algorithm used to solve the relaxations of \ref{Eq:1maxy} and \ref{Eq:3min0}. They are based on a {\em proximal augmented Lagrangian (AL)} method in which, at each round, the associated unconstrained subproblem is solved in a federated manner by an inexact {\em alternating direction method of multipliers (ADMM)}. We adopt this framework from~\cite{He2024FederatedConstraints}, whose exposition is included here for completeness (see \cref{sec:prox-al} and \cref{sec:admm} below). Figure~\ref{fig3} shows the structure of the FL framework.

\subsubsection{A proximal AL based FL algorithm for solving \cref{eq:FL_Framework}}\label{sec:prox-al}
In this subsection, we describe a proximal AL-based FL algorithm (\cref{alg:g-admm-cvx}) for finding a (nearly) optimal solution of \cref{eq:FL_Framework} for prescribed $\epsilon_1,\epsilon_2\in(0,1)$. For notational convenience, we omit the data symbols $Z_m$ in local objectives and constraints when no confusion arises, and we denote by $\mu_m$ the Lagrangian multiplier associated with constraint $c_m$ with $m=0$ referring to the global constraint. Each $c_m(\phi)$ is understood as the vector formed by stacking the corresponding scalar local or global inequality constraints, we denote by \(p_m\) the number of stacked scalar inequalities, i.e., the dimension of \(c_m(\phi)\). The operator $[\cdot]_+$ acts elementwise. The multiplier \(\mu_m\in\mathbb{R}_+^{p_m}\) is the nonnegative vector with the same dimension as \(c_m(\phi)\). At outer round \(k\), the proximal AL subproblem associated with \cref{eq:FL_Framework} is defined as

\begin{equation} \label{prox-AL-pb}
\min_{\phi}\Bigg\{\ell_k(\phi):= \underbrace{\sum_{m=1}^N f_m(\phi) + h(\phi) + \frac{1}{2\beta}\sum_{m=0}^N\left(\|[\mu_m^k+\beta c_m(\phi)]_+\|^2-\|\mu_m^k\|^2\right)}_{\text{augmented Lagrangian function}} + \underbrace{\frac{1}{2\beta}\|\phi-\phi^k\|^2}_{\text{proximal term}}\Bigg\}.
\end{equation}

Then, the multiplier estimates are updated according to the classical scheme:

\begin{equation}
\mu_m^{k+1} = [\mu^k_m+\beta c_m(\phi^{k+1})]_+,\quad 0\le m\le N.
\end{equation}

The method uses the sequential penalization approach with parameter $\beta$, which involves solving a sequence of unconstrained subproblems that combine the objective function with penalization of constraint violations.

\begin{algorithm}[h]
\caption{A proximal AL based FL algorithm}
\label{alg:g-admm-cvx}
\noindent\textbf{Input}: tolerances $\epsilon_1,\epsilon_2\in(0,1)$, $\phi^0$, $\mu_m^0\ge0$ for $0\le m\le N$, $\Bar{s}>0$, and $\beta>0$.
\begin{algorithmic}[1]
\For{$k=0,1,2,\ldots$}
\State Set $\delta_k=\Bar{s}/(k+1)^2$.\label{def:beg-outer-loop}
\State Call \cref{alg:admm-cvx-1} (see \cref{sec:admm} below) with $(\delta,\tilde{\phi}^0)=(\delta_k,\phi^k)$ to find an approximate solution $\phi^{k+1}$ 
\State to \cref{prox-AL-pb-compact} in a federated manner such that
\begin{equation}\label{cond:apx-stat}
\mathrm{dist}_{\infty}(0,\partial\ell_k(\phi^{k+1}))\le\delta_k.
\end{equation}
\State {\bf Coordinator update:} The central coordinator updates $\mu_0^{k+1}=[\mu_0^k + \beta c_0(\phi^{k+1})]_+$.
\State {\bf Communication (broadcast):} Each local DT $m$, $1\le m\le N$, receives $\phi^{k+1}$ from the coordinator.
\State {\bf DT update (local):} Each local DT $m$, $1\le m\le N$, updates $\mu_m^{k+1}=[\mu_m^k + \beta c_m(\phi^{k+1})]_+$.
\State {\bf Communication:} Each local DT $m$, $1\le m\le N$, sends $\|\mu_m^{k+1}-\mu_m^k\|_\infty$ to the coordinator.
\State {\bf Termination (Coordinator side):} Output $(\phi^{k+1},\mu^{k+1})$ and terminate the algorithm if
\begin{equation}\label{stop-alg}
\|\phi^{k+1}-\phi^k\|_{\infty}+\beta\delta_k\le\beta\epsilon_1,\qquad \max_{0\le m\le N}\{\|\mu_m^{k+1}-\mu_m^k\|_{\infty}\}\le\beta\epsilon_2.
\end{equation} 
\label{def:end-outer-loop}
\EndFor
\end{algorithmic}
\end{algorithm}

Notice that the subproblem in \cref{prox-AL-pb} can be rewritten as 
\begin{equation}\label{prox-AL-pb-compact}
\min_\phi\left\{\ell_k(\phi) := \sum_{m=0}^N P_{m,k}(\phi) + h (\phi)\right\}, 
\end{equation}
where $P_{m,k}$, $0\le m\le N$, are defined as
\begin{align}
P_{0,k}(\phi):= &\ \frac{1}{2\beta}\left(\|[\mu_0^k+\beta c_0(\mathbf{\phi})]_+\|^2-\|\mu_0^k\|^2\right) + \frac{1}{2(N+1)\beta}\|\phi-\phi^k\|^2,\label{def:P0k}\\
P_{m,k}(\phi):= &\ f_m(\phi) + \frac{1}{2\beta}\left(\|[\mu_m^k+\beta c_m(\phi)]_+\|^2-\|\mu_m^k\|^2\right) + \frac{1}{2(N+1)\beta}\|\phi-\phi^k\|^2,\quad \forall 1\le m\le N.\label{def:Pik}
\end{align}

At each round of \cref{alg:g-admm-cvx}, it applies an inexact ADMM solver (\cref{alg:admm-cvx-1}) to find an approximate solution \(\phi^{k+1}\) to the subproblem \cref{prox-AL-pb-compact}, in which the local merit function \(P_{m,k}\), constructed from the local objective $f_m$ and local constraint $c_m$, is handled by the respective local client \(m\), while the merit function $P_{0,k}$ is handled by the central coordinator.

For ease of later reference, we refer to the update from $\phi^k$ to $\phi^{k+1}$ as one round of \cref{alg:g-admm-cvx}, and call one iteration of \cref{alg:admm-cvx-1} for solving \cref{prox-AL-pb} one inner iteration of \cref{alg:g-admm-cvx}.

\subsubsection{An inexact ADMM for FL}\label{sec:admm}

\begin{algorithm}[h]
\caption{An inexact ADMM based FL algorithm}
\label{alg:admm-cvx-1}
\noindent\textbf{Input}: tolerance $\delta\in(0,1]$, $q\in(0,1)$, $\tilde{\phi}^0$, and $\rho_m>0$ for $1\le m\le N$;
\begin{algorithmic}[1]
\State Set $\phi^0=\tilde{\phi}^0$, and $(u_m^0,\lambda_m^0,\tilde{u}_m^0)=(\tilde{\phi}^0,-\nabla P_m(\tilde{\phi}^0),\tilde{\phi}^0-\nabla P_m(\tilde{\phi}^0)/\rho_m)$ for $1\le m\le N$.
\For{$t=0,1,2,\ldots$}
\State Set $\varepsilon_{t+1}=q^t$;\label{def:for-loop-beg}
\State {\bf Coordinator update:} The central server finds an approximate solution $\phi^{t+1}$ to
\begin{equation}\label{sbpb:phi-gt}
\min_\phi\left\{\varphi_{0,t}(\phi):=P_0(\phi)+h(\phi)+\sum_{m=1}^N\left[\frac{\rho_m}{2}\|\tilde{u}_m^t-\phi\|^2\right]\right\}
\end{equation} 
\State such that $\rmdist_{\infty}(0,\partial \varphi_{0,t}(\phi^{t+1}))\le\varepsilon_{t+1}$.
\State {\bf Communication (broadcast):}  Each local DT $m$, $1\le m\le N$, receives $\phi^{t+1}$ from the coordinator. 
\State {\bf Client update (local):} Each local DT $m$, $1\le m\le N$, finds an approximate solution $u_m^{t+1}$ to
\begin{equation}\label{sbpb:phi-it}
\min_{u_m}\left\{\varphi_{m,t}(u_m) := P_m(u_m) + \langle\lambda_m^t,u_m-\phi^{t+1}\rangle + \frac{\rho_m}{2}\|u_m-\phi^{t+1}\|^2\right\}
\end{equation}
\State such that $\|\nabla \varphi_{m,t}(u_m^{t+1})\|_{\infty}\le\varepsilon_{t+1}$, and then updates
\begin{align}
\lambda_m^{t+1}   = &\ \lambda_m^t + \rho_m(u_m^{t+1}-\phi^{t+1}),\label{sbpb:lam-gt}\\
\tilde{u}_m^{t+1} = &\ u_m^{t+1}+\lambda_m^{t+1}/\rho_m,\label{def:tuit}\\
\tilde{\varepsilon}_{m,t+1} = &\ \|\nabla\varphi_{m,t}(\phi^{t+1})-\rho_{m}(\phi^{t+1}-u_m^t)\|_{\infty}.\label{vareps-gt}
\end{align}
\State {\bf Communication:} Each local DT $m$, $1\le m\le N$, sends $(\tilde{u}_m^{t+1},\tilde{\varepsilon}_{m,t+1})$ back to the coordinator.
\State {\bf Termination (Coordinator side):} Output $\phi^{t+1}$ and terminate this algorithm if
\begin{equation}\label{sbpb:stop}
\varepsilon_{t+1} + \sum_{m=1}^N\tilde\varepsilon_{m,t+1}\le \delta.
\end{equation}\label{def:for-loop-end}
\EndFor
\end{algorithmic}
\end{algorithm}

\begin{figure*}[t!]
    \centering
    \includegraphics[width=\textwidth]{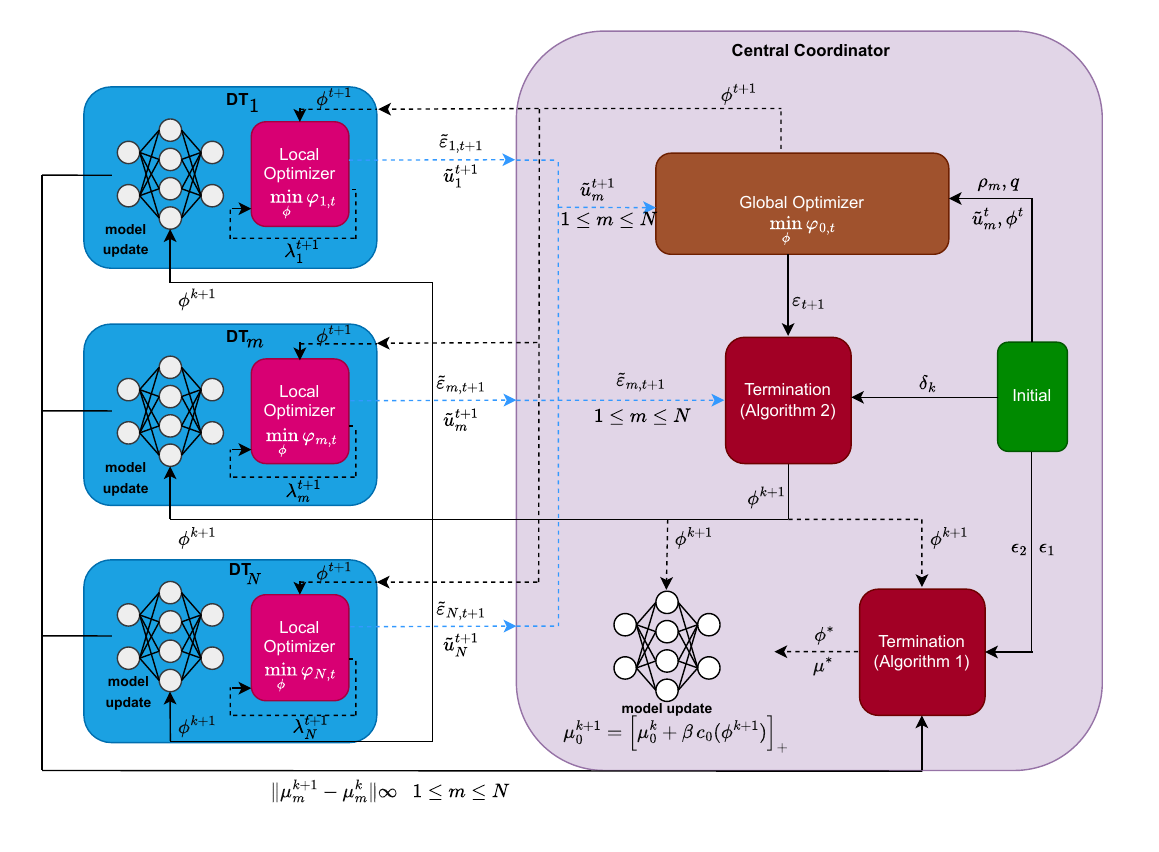}
   	\caption{Structure of proximal AL-based FL algorithm.}
   	\label{fig3}
\end{figure*}

In this subsection, we describe an inexact ADMM-based FL algorithm (\cref{alg:admm-cvx-1}) for solving \cref{prox-AL-pb-compact}. We need to solve the subproblem of the same form  each $k$, therefore we drop $k$ and focus on solving the following model problem in an FL  manner. To make each participating client $m$ handle their local objective $P_m$ independently (see \cref{sec:prox-al}), we introduce decoupling variables $u_m$'s and obtain the following equivalent consensus reformulation for \cref{prox-AL-pb-compact}: 
\begin{equation}\label{pb:consens}
\min_{\phi,u_m} \left\{\sum_{m=1}^N P_m(u_m) + P_0(\phi) + h(\phi)\right\}\quad \st\quad u_m=\phi,\quad 1\le m\le N,
\end{equation}
which allows each local client $m$ to handle the local variable $u_m$ and the local objective function $P_m$ while imposing consensus constraints that force clients’ local parameters $u_m$ equal to the global parameter $\phi$. This reformulation enables the applicability of an inexact ADMM that solves \cref{pb:consens} in a federated manner. At each iteration, an ADMM solver optimizes the AL function associated with \cref{pb:consens}:
\begin{equation}\label{AL-consensus-F}
\cL_P(\phi,u,\lambda):=\sum_{m=1}^N \left[P_m(u_m)+\langle\lambda_m, u_m-\phi\rangle + \frac{\rho_m}{2}\|u_m-\phi\|^2\right] + P_0(\phi) + h(\phi)
\end{equation}
with respect to the variables $\phi$, $u$, and $\lambda$ alternately, where $u=[u_1^T,\ldots,u_N^T]^T$ and $[\lambda_1^T,\ldots,\lambda_N^T]^T$ collect all the local parameters and the multipliers associated with the consensus constraints, respectively. Specifically, in iteration $t$, one performs
\begin{align}
&\phi^{t+1}\approx\argmin_\phi \cL_P(\phi,u^t,\lambda^t),\label{w-admm}\\
&u^{t+1}\approx\argmin_{u} \cL_P(\phi^{t+1},u,\lambda^t),\label{u-admm}\\
&\lambda_m^{t+1} = \lambda_m^t + \rho_m (u_m^{t+1}-\phi^{t+1}),\quad \forall 1\le m\le N.
\end{align}
By the definition of $\cL_P$ in \cref{AL-consensus-F}, one can verify that the step in \cref{w-admm} is equivalent to \cref{sbpb:phi-gt}, and also the step in \cref{u-admm} can be computed in parallel, which corresponds to \cref{sbpb:phi-it}. Therefore, the ADMM updates naturally suit the FL framework, as the separable structure in \cref{AL-consensus-F} over the pairs $\{(u_m,\lambda_m)\}$ enables the local update of $(u_m,\lambda_m)$ at each client $m$ while $\phi$ is updated by the central coordinator.

By setting $B{\log _2}( 1 + \frac{{\bar{P}_{m}}{g_{m,t}}}{N_0Bx})x|_{x=0}:=0$, Assumptions 0 and 1 of~\cite{He2024FederatedConstraints} (convexity and continuous differentiability of objectives and constraints, and strong duality for \ref{Eq:1maxy} and \ref{Eq:3min0}) are satisfied. Unfortunately the Lipschitz conditions of Assumption 2 (b) hold only for \ref{Eq:3min0} (while Assumption 2 (a) is trivially true for both problems), and this will affect our complexity analysis for \ref{Eq:1maxy} below.

By applying the FL algorithm of \cite{He2024FederatedConstraints} we obtain a (nearly) optimal fractional solution for \ref{Eq:1maxy}, \ref{Eq:3min0} (Theorem 7 in \cite{He2024FederatedConstraints}), without disclosing private information, as guaranteed by FL.

Note that one could argue that $n_m$ (for \ref{Eq:1maxy}) or $\lceil C_{m}/(R\tau)\rceil$ (for \ref{Eq:3min0}) can be deduced by the (globally available) transmission schedule $\mathbf{x}$, since the latter discloses how many time slots each user uses. But determining exactly these values, and not a lower bound for them, is not possible since only user $m$ knows whether the fractional solution makes constraint \cref{Eq:1C3} or \cref{Eq:3C3} tight or not.

\subsection{Rounding the fractional solution}  
\label{sub:round}

In what follows, we describe our algorithms used to round $x_f$ to an integral schedule $x^*$. They are based on {\em dependent rounding} \cite{Gandhi2006}, which we employ similarly to~\cite{Chen2024}, and whose exposition is included here for completeness. Assume we are given a bipartite graph $(V_1, V_2, E)$ with bipartition $(V_1, V_2)$ and a value $x_{i,j} \in [0, 1]$ for each edge $(i, j) \in E$. Initialize $y_{i,j} = x_{i,j}$ for each $(i, j) \in E$. Values $y_{i,j}$ will be probabilistically modified in several (at most $|E|$) iterations such that $y_{i,j} \in \{0, 1\}$ at the end, at which point we will set $x_{i,j} := y_{i,j}$ for all $(i, j) \in E$, where now
$x_{i,j}$ are the (random) rounded final values for edges $(i,j)\in E$.

The iterations that modify $y$ proceed as follows: We call an edge $(i,j)$ {\em floating} if its value $y_{i,j}$ is not integral (i.e., $y_{i,j}\in(0,1)$). Let $\tilde E \subseteq E$ be the current set of floating edges. If $\tilde E = \emptyset$, the process terminates by setting $x_{i,j} := y_{i,j}$ for all $(i, j) \in E$. Otherwise, find a maximal path $S$ in the subgraph $(V_1, V_2, \tilde E)$ in $O(|V_1|+|V_2|)$ time running depth-first-search (DFS). Partition the edge-set of $S$ into two alternating matchings $A$ and $B$. We define
\begin{equation}\label{Eq:alpha}
\alpha:=\min\{\gamma >0: (\exists(i,j) \in A : y_{i,j}+ \gamma =1) \ \vee\ (\exists(i,j) \in B : y_{i,j}- \gamma =0)\}
\end{equation}

\begin{equation}\label{Eq:beta}
\beta:=\min\{\gamma >0: (\exists(i,j) \in A : y_{i,j}- \gamma =0) \ \vee\ (\exists(i,j) \in B : y_{i,j}+ \gamma =1)\}.
\end{equation}
Then we execute the following randomized step:
\begin{itemize}
\item With probability $\beta/(\alpha+\beta)$ set $y_{i,j} := y_{i,j} + \alpha\ \forall (i,j)\in A, y_{i,j} := y_{i,j} - \alpha\ \forall (i,j)\in B$, and
\item with probability $\alpha/(\alpha+\beta)$ set $y_{i,j} := y_{i,j} - \beta\ \forall (i,j)\in A, y_{i,j} := y_{i,j} + \beta\ \forall (i,j)\in B$.
\end{itemize}
In either case, at least one edge $(i,j)\in\tilde E$ will stop being floating, i.e., $y_{i,j}\in\{0,1\}$, and, therefore, after at most $|E|$ iterations or $O(|E|(|V_1|+|V_2|))$ time, all values $y$ will become integral and the rounding process terminates. We refer to this algorithm as {\bf Simple Dependent Rounding}, which is summarized in \cref{alg:dp}. Note that it is a randomized algorithm, and the final rounded solutions $x^*$ are random variables. In the context of our problem,
$V_1=\{m:\, m\in\mathcal{N}\}$, $V_2=\{t:\, t\in\mathcal{T}\}$, and $E=\{(m,t): x_{m,t}>0\}$. Also, since dependent rounding never violates the degree of the nodes of $G$, constraints \cref{Eq:1C3} or \cref{Eq:3C3} are satisfied from the beginning to the end of the algorithm when applied to the fractional solution of \ref{Eq:1maxy} or \ref{Eq:3min0}, respectively. 

\begin{algorithm}[h]
\caption{Simple Dependent Rounding}
\label{alg:dp}
\noindent\textbf{Input}: bipartite graph $G=(V_1,V_2,E)$; fractional assignments $x_{i,j}\in[0,1]$ for all $(i,j)\in E$.
\begin{algorithmic}[1]
\State Initialize $y_{i,j}\gets x_{i,j}$ for all $(i,j)\in E$; \ $\tilde E\gets E$.
\ForAll{$(i,j)\in \tilde E$}
  \If{$y_{i,j}\in\{0,1\}$}
    \State $\tilde E\gets \tilde E\setminus\{(i,j)\}$
  \EndIf
\EndFor
\While{$\tilde E\neq\emptyset$}
  \State \textbf{Path selection:} Find a maximal alternating path $S\gets\mathrm{DFS}(V_1,V_2,\tilde E)$; let $S=A\cup B$ with alternating matchings $A,B$.
  \State \textbf{Step sizes:} Compute $\alpha$ and $\beta$ from \cref{Eq:alpha} and \cref{Eq:beta}.
  \State With probability $\beta/(\alpha+\beta)$:
  \Statex \quad $y_{i,j}\gets y_{i,j}+\alpha$ for all $(i,j)\in A$, and $y_{i,j}\gets y_{i,j}-\alpha$ for all $(i,j)\in B$.
  \State With probability $\alpha/(\alpha+\beta)$:
  \Statex \quad $y_{i,j}\gets y_{i,j}-\beta$ for all $(i,j)\in A$, and $y_{i,j}\gets y_{i,j}+\beta$ for all $(i,j)\in B$.
  \ForAll{$(i,j)\in \tilde E$}
    \If{$y_{i,j}\in\{0,1\}$}
      \State $\tilde E\gets \tilde E\setminus\{(i,j)\}$
    \EndIf
  \EndFor
\EndWhile
\State Set $x_{i,j}\gets y_{i,j}$ for all $(i,j)\in E$.
\State \Return $x$
\end{algorithmic}
\end{algorithm}

We propose a modification of this standard dependent rounding algorithm (called {\bf Smart Dependent Rounding}), that uses a limited amount of information about the users' channel gain factors $g_{m,t}$ in the hope of achieving a better rounding. Namely, dependent rounding is applied separately first on the set of edges with large link gains $g_{m,t}\geq \theta$ for some threshold $\theta$, then on the set of edges with link gains $g_{m,t}<\theta$, and, finally, on all remaining floating edges so that in the end an integral schedule is calculated. A step-by-step description is given in \cref{alg:smart-dp}. Note that in lines \ref{line:firstapply}-\ref{line:thirdapply} each edge value $y_{i,j}$ that is rounded to $0$ or $1$ is not a floating edge for subsequent applications of \cref{alg:dp}. The intuition behind the prioritization of large link gain edges for rounding is that the fractional solution would give large value $x_{i,j}$ to such edges $(i,j)$ for users that are tightly constrained by their demands. Hence, in order to satisfy \cref{Eq:1C3} or \cref{Eq:3C3} they should be scheduled on high-gain links for transmission.

More specifically, and since link gains are private, $\theta$ is computed by the central coordinator, in collaboration with the user DTs, using a Gaussian Mixture Model (GMM) approach that doesn't reveal the $g_{m,t}$'s. In the case of Jakes's model~\cite{jakes1994microwave}, a two-component GMM is fit in a federated EM manner~\cite{Dieuleveut2021FedEM} to the observed link gains, preserving the privacy of the individual $g_{m,t}$ since no raw data leaves the local users (only aggregated sufficient statistics are shared), and $\theta$ is determined as the intersection point of the resulting Gaussian distributions. This GMM-based thresholding provides a lightweight, data-driven way to separate weak/strong gain regimes and set $\theta$ without centralizing raw link gains. It is important to emphasize that the proposed Smart Dependent Rounding framework is independent of both the channel model and the specific method used to determine $\theta$. After computing $\theta$, the central coordinator is asking each DT $m$ to return $0$ if $g_{m,t}<\theta$ or $1$ otherwise (this is the only bit of information about gain factors that users need to reveal). Then the edge grouping and the dependent rounding can proceed as described above.

\begin{algorithm}[h]
\caption{Smart Dependent Rounding}
\label{alg:smart-dp}
\noindent\textbf{Input}: $G=(V_1,V_2,E)$, $x_{i,j}\in[0,1]$ for each $(i,j)\in E$; $[g_{m,t}]_{t=1}^T$ for all $m$.
\begin{algorithmic}[1]
\State \textbf{Preprocessing (Coordinator):} In collaboration with the DTs, estimate $\theta$ via a two-component GMM fitted with federated EM~\cite{Dieuleveut2021FedEM}. \textbf{Broadcast} $\theta$ to all DTs. \label{line:broadcast-theta}

\State \textbf{Local step (each DT $m$):} For each owned edge $(m,t)$, compute
\[
b_{m,t}=\begin{cases}
1, & g_{m,t}\ge \theta \ \\
0, & g_{m,t}< \theta \ 
\end{cases}
\]
\State \textbf{Communication:} Each DT sends only $\,b_{m,t}$ (never $g_{m,t}$) to the coordinator.
\State \textbf{Coordinator initialization:} $y_{i,j}\gets x_{i,j}$ for all $(i,j)\in E$.
\State \textbf{Coordinator partition:} 
\[
E_1:=\{(m,t)\in E:\ b_{m,t}=1\},\qquad
E_2:=\{(m,t)\in E:\ b_{m,t}=0\}.
\]
\State \textbf{Coordinator:} Apply \cref{alg:dp} on graph $G_1=(V_1,V_2,E_1)$. \phantomsection\label{line:firstapply}
\State \textbf{Coordinator:} Apply \cref{alg:dp} on graph $G_2=(V_1,V_2,E_2)$. \phantomsection\label{line:secondapply}
\State \textbf{Coordinator:} Apply \cref{alg:dp} on graph $G=(V_1,V_2,E)$. \phantomsection\label{line:thirdapply}
\State Set $x_{i,j}\gets y_{i,j}$ for all $(i,j)\in E$.
\State \Return $x$
\end{algorithmic}
\end{algorithm}    

\cref{alg:fp} ({\bf Fixed-Power Scheduler (FPS)}) below codifies the solution steps for problem P1 (problem P2 is similar), while \cref{alg:pc} ({\bf Power-Control Scheduler (PCS)}) codifies the solution steps for problem P3. Lines~\ref{a1:l1} in both algorithms apply binary search on $T$ in order to determine the minimum $T_f$ for which the FL algorithm of~\cite{He2024FederatedConstraints} produces a feasible solution. Lines~\ref{a1:l2}-\ref{a1:l6} round $x_f$ using either the {\bf Simple Dependent Rounding} or the {\bf Smart Dependent Rounding} (depending on choice \texttt{rounding\_mode}). \cref{alg:fp} applies an extra step of re-optimization for bandwidth allocation $w$ (line~\ref{a1:l7}) to satisfy as many of the \cref{Eq:1C2} constraints as possible, applying again the algorithm of~\cite{He2024FederatedConstraints}. Finally, if there are still unsatisfied constraints \cref{Eq:1C2}, \cref{alg:fp} increases (global) bandwidth allocation $B$ (line~\ref{a1:l8}) by satisfying user requests. In \cref{alg:pc}, if there are unsatisfied constraints \cref{Eq:3C2}, users increase their energy allocations $E_m$ (line~\ref{a2:l7}) until these constraints are satisfied. Note that this step is done internally by each user, so no private data are revealed.  

\begin{algorithm}[h]
\caption{Fixed-Power Scheduler (FPS) for P1}
\label{alg:fp}
\noindent\textbf{Input}: user data $C_m, n_m, \bar P_m, [g_{m,t}]_{t=1}^T$ for all $m$; system parameters $B, N_0, \tau$.
\begin{algorithmic}[1]
\State \textbf{Binary search on $T$}: obtain $x_f, w_f, \gamma_f, T_f$. \phantomsection\label{a1:l1}
\If{\texttt{rounding\_mode} $==$ \texttt{simple}} \phantomsection\label{a1:l2}
  \State \textbf{Simple Dependent Rounding}$(x_f, w_f) \rightarrow x_r, w_r$
\Else
  \State \textbf{Smart Dependent Rounding}$(x_f, w_f) \rightarrow x_r, w_r$
\EndIf \phantomsection\label{a1:l6}
\State \textbf{Re-optimization}$(x_r, T_f) \rightarrow w_{ro}$ \phantomsection\label{a1:l7}
\State \textbf{Boost bandwidth}$(x_r, w_{ro}, B) \rightarrow w^{*}, B^{*}$ \phantomsection\label{a1:l8}
\State \Return $T_f, x_r, w^{*}, B^{*}$
\end{algorithmic}
\end{algorithm}

\begin{algorithm}[h]
\caption{Power-Control Scheduler (PCS) for P3}
\label{alg:pc}
\noindent\textbf{Input}: user data $C_m, E_m, [P_{m,t}]_{t=1}^T$ for all $m$; system parameters $R, \tau$.
\begin{algorithmic}[1]
\State \textbf{Binary search on $T$}: obtain $x_f, T_f$.
\If{\texttt{rounding\_mode} $==$ \texttt{simple}} \phantomsection\label{a2:l2}
  \State \textbf{Simple Dependent Rounding}$(x_f) \rightarrow x_r$
\Else
  \State \textbf{Smart Dependent Rounding}$(x_f) \rightarrow x_r$
\EndIf
\State \textbf{Boost energy}$(x_r)$ \phantomsection\label{a2:l7}
\State \Return $T_f, x_r$
\end{algorithmic}
\end{algorithm}
    
\subsection{Transmission periodicity}
\label{subsec:periodicity}

Besides ensuring the security of private data, the proposed solution must also comply with the requirement of periodic transmission, a feature of applications using DTs. To achieve this, we have to accelerate the computation of the next-round transmission schedule and bandwidth allocation (for problems P1 and P2), so that it can be completed {\em before} the current transmission round is completed.

We start by noting that the binary search employed for the solution of problems P1 and P3 takes at most $O(\log T_f)$ iterations. In every iteration, the FL algorithm of \cite{He2024FederatedConstraints} runs in iterations, each computing a better primal or dual solution of a convex problem, related to relaxations \ref{Eq:1maxy}-\ref{Eq:3min0}, and each such problem can be solved by a standard solver in polynomial time on $N$ and $T_f$. Since the (linear) constraints of \ref{Eq:3min0} satisfy the Lipschitz conditions of Assumption 2(b) of \cite{He2024FederatedConstraints}, the number of iterations is $O(\max\{\epsilon_1^{-4}, \epsilon_2^{-4}\})$, where $\epsilon_1, \epsilon_2$ are implementation parameters that determine the maximum deviation from the optimal solution and satisfaction of constraints, respectively, in the sense of $\epsilon$-KKT approximability (Theorem 4 of \cite{He2024FederatedConstraints}). Unfortunately, constraints \cref{Eq:1C2} of \ref{Eq:1maxy} and \cref{Eq:2C2} of \ref{Eq:2maxy} do not satisfy these conditions. {In particular, Assumption~2(b) of~\cite{He2024FederatedConstraints} requires the Jacobians of the local constraints $c_m(\phi; Z_m)$ (which include \cref{Eq:1C2} and \cref{Eq:2C2}) to be locally Lipschitz, i.e.,
\begin{equation}
\label{eq:lipschitz_local_inline}
\|\nabla \phi(u)-\nabla \phi(v)\|\le L \|u-v\|,\qquad \forall\, u,v\in\mathcal{U},
\end{equation}
for some neighborhood $\mathcal{U}$ and constant $L>0$. 
In \cref{Eq:1C2} and \cref{Eq:2C2},
\[ \phi_{m,t}(w)\triangleq B\,w\,\log_2\!\left(1+\frac{\alpha_{m,t}}{w}\right),\qquad \alpha_{m,t}\triangleq \frac{\bar{P}_m g_{m,t}}{N_0B}>0. \]
It is easy to see that this function violates \cref{eq:lipschitz_local_inline} at $w=0$: its derivative satisfies $\phi'_{m,t}(w)\to +\infty$ as $w\to 0^+$, and, therefore, it is not locally Lipschitz around $w=0$. In order to avoid iterating a lot when $\phi_{m,t}(w)\rightarrow 0^+$, i.e., when $w_{m,t}\rightarrow 0$, we force bandwidth allocation $w_{m,t}:=0$. Although we have no theoretical upper bound for the number of iterations in this case, the number of iterations is greatly decreased by this heuristic in our simulations (\cref{sec:numerical}), and this fast execution allows for the repeated application of our scheduling algorithms within small time periods. The downside is that the heuristic for handling the violated Lipschitz conditions may introduce additional approximation error.} 

Finally, the rounding algorithms need $O(N^2T^2_f)$ time.

\section{Numerical Results}
\label{sec:numerical}

In this section, we present simulation results to demonstrate the performance of the proposed solution. We consider a single BS serving $N$ users. The initial positions of the users are uniformly distributed within a circular area of radius 1\,km centered at the BS. Each user moves at a constant speed $v$, and their moving directions are independently and uniformly distributed over $[0,2\pi]$. Let $d_{m,t}$ denote the distance between user $m$ and the BS at time $t$ and $L_{m,t} = L_0 (d_{m,t})^{-\alpha}$ the distance-based path loss, where $L_0$ is the path loss when the distance is 1\,m and $\alpha$ is the path loss exponent. The complex channel gains between a user and the BS are generated from a first-order Gauss–Markov process~\cite{Sklar2001} as follows:
\begin{equation}
  h_{m,t+1} = \rho\,h_{m,t} + \sqrt{1-|\rho|^2}\,\omega_{m,t},
\end{equation}
where $\omega_{m,t} \sim \mathcal{CN}\!\big(0,\,L_{m,t}\big)$ is the complex Gaussian noise with zero mean and variance $L_{m,t}$, and
$\rho = J_0(2\pi f_d \tau)$ is the correlation coefficient between two successive time slots, in accordance with Jakes\textquotesingle{} model~\cite{Sklar2001}. Here, $J_0(\cdot)$ is the zeroth-order Bessel function, $f_d = \tfrac{v f_c}{c}$ is the Doppler shift, $f_c$ is the carrier frequency, and $c$ is the electromagnetic wave propagation speed.
The channel power gain between user $m$ and the BS at time slot $t$ is then
\begin{equation}
  g_{m,t} = |h_{m,t}|^2 .
\end{equation}

Consistent with our assumption that either the number of users requiring simultaneous scheduling is small or that users can be partitioned into small, independent groups, we set the number of users to $N=10$, which ensures that the runtime of our algorithms remains well below the scheduling time period. In addition, the scheduling horizon is set to $T=100$ time slots, reflecting the practical limitation that reliable future channel state knowledge or prediction is typically unavailable beyond a short time horizon.

Default parameters used in the simulation are summarized in \cref{tab:parameters}, whereas the algorithm-specific hyperparameters of the proposed FL framework are listed in \cref{tab:fl-framework-parameters}. Similar parameter values were also used in \cite{mcmahan2017communication,Chen2024,Kim2022}.

The hyperparameters of the federated optimization framework are selected to balance constraint satisfaction, numerical stability, and execution latency. The augmented--Lagrangian penalty $\beta$ is chosen so that constraint violations are sufficiently penalized without causing ill-conditioning in the local convex subproblems. The local step size $\rho_m$ is set conservatively (i.e., it is small) to ensure stable client/DT updates under heterogeneous channel and demand conditions. A small proximal parameter $\bar{s}$ is used to regularize the local updates and improve robustness near boundary solutions (e.g., very small allocations). The primal and dual tolerances $(\varepsilon_1,\varepsilon_2)$ are selected to achieve reliable convergence with negligible practical constraint violations while keeping the number of iterations small. We use a relaxation/momentum factor $q$ close to one to mildly accelerate convergence without inducing oscillations. Finally, the local multipliers $\mu_m^0$ are initialized to zero to provide a neutral starting point when no prior dual information is available. The specific values of these hyperparameters in \cref{tab:fl-framework-parameters} were finalized through trial-and-error tuning to obtain stable convergence and good performance.

The hyperparameters of the federated GMM used to determine $\theta$ were finalized through trial-and-error tuning to achieve stable clustering and good performance on the Jakes-model-based channel traces employed in our simulations.

The simulations were performed on a desktop with an Intel\textsuperscript{\textregistered} Core i7-12700 CPU (12~cores/20~threads, base~2.10\,GHz), 64\,GB RAM; the system ran a 64-bit operating system.

To evaluate our proposed algorithms, we compared their performance against Bonmin~\cite{bonmin}, an open-source MINLP solver from the COIN-OR project~\cite{bonami2008algorithmic}, licensed under the Eclipse Public License.


\begin{table}[t]
  \centering
  \caption{Default simulation parameters.}
  \label{tab:parameters}
  \vspace{2pt}
  \footnotesize
  \setlength{\tabcolsep}{6pt} 
  \begin{tabular}{@{}ll@{}}
    \cmidrule(l{-6pt}r{-6pt}){1-2}
    \textbf{Parameter} & \textbf{Value} \\
    \cmidrule(l{-6pt}r{-6pt}){1-2}
    Number of users $N$                         & $10$ \\
    Slot duration $\tau$                        & $5$ ms \\
    Update period $T$                           & $100$ slots \\
    Block length (for P2) $t_s$                 & $10$ slots \\
    Per-user demand $C_m$                       & $1.6$ Mbit \\
    Fixed transmit power $\bar{P}_m$            & $0.1$ W \\
    User speed $v$                              & $10$ km/h \\
    System bandwidth $B$                        & $9$ MHz \\
    Noise spectral density $N_0$                & $-174$ dBm/Hz \\
    Fixed uplink rate $R$ (single-user slotting)& $80$ Mb/s \\
    Energy budget $E_m$                         & $40$ mJ \\
    Carrier frequency $f_c$                     & $900$ MHz \\
    Path loss exponent                          & $3$ \\
    \cmidrule(l{-6pt}r{-6pt}){1-2}
  \end{tabular}
  \vspace{-2mm}
\end{table}


\begin{table}[t]
  \centering
  \caption{Algorithm-specific hyperparameters for the federated
optimization/learning framework.}
  \label{tab:fl-framework-parameters}
  \vspace{2pt}
  \footnotesize
  \setlength{\tabcolsep}{6pt}
  \renewcommand{\arraystretch}{1.1}
  \begin{tabular}{@{}l c@{}}
    \cmidrule(l{-6pt}r{-6pt}){1-2}
    \textbf{Parameter} & \textbf{Value} \\
    \cmidrule(l{-6pt}r{-6pt}){1-2}
    Augmented–Lagrangian penalty $\beta$               & $15$ \\
    Local step size $\rho_m$                           & $0.01$ \\
    Proximal parameter $\bar{s}$                       & $0.001$ \\
    Primal tolerance $\epsilon_{1}$                 & $0.5$ \\
    Dual tolerance $\epsilon_{2}$                   & $0.05$ \\
    Relaxation/momentum $q$                            & $0.9$ \\
    Initial local multipliers $\mu_m^{0}$ & $(0,\ldots,0)^{\mathsf T}$ \\[1pt]
        \cmidrule(l{-6pt}r{-6pt}){1-2}
  \end{tabular}
  \vspace{-2mm}
\end{table}

\subsection{Simulation Set I: resource allocation}

\begin{figure}[H]
  \centering
  \subfloat[P1: boost vs. demand\label{set-one-P1-demand}]{
    \includegraphics[width=0.47\columnwidth]{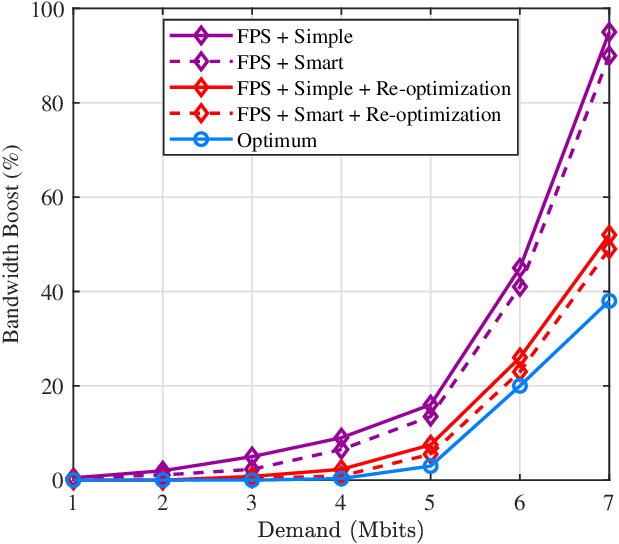}}\hfill
  \subfloat[P1: boost vs. bandwidth\label{set-one-P1-bandwidth}]{
    \includegraphics[width=0.47\columnwidth]{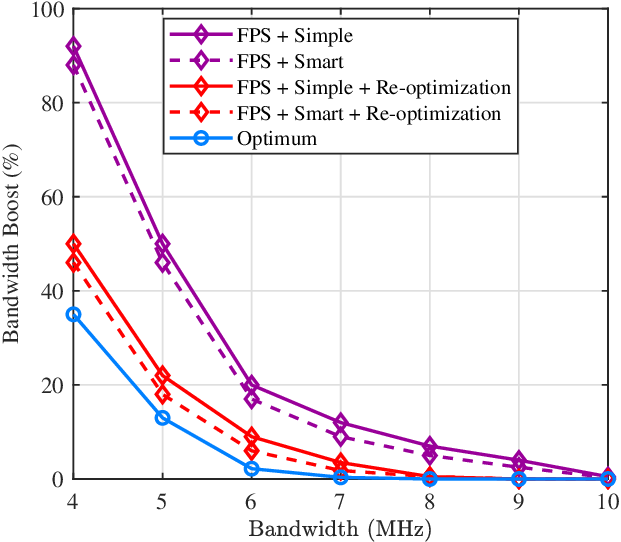}}
  \caption{Simulation Set I (FPS, P1): required bandwidth boost under varying demand $C_m$ and bandwidth $B$.}
  \label{fig:set1_fps_p1}
\end{figure}

\begin{figure}[H]
  \centering
  \subfloat[P2: boost vs. demand\label{set-one-P2-demand}]{
    \includegraphics[width=0.47\columnwidth]{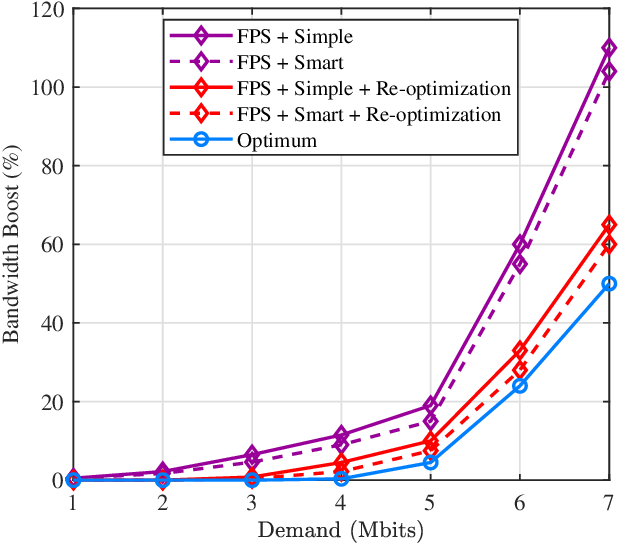}}\hfill
  \subfloat[P2: boost vs. bandwidth\label{set-one-P2-bandwidth}]{
    \includegraphics[width=0.47\columnwidth]{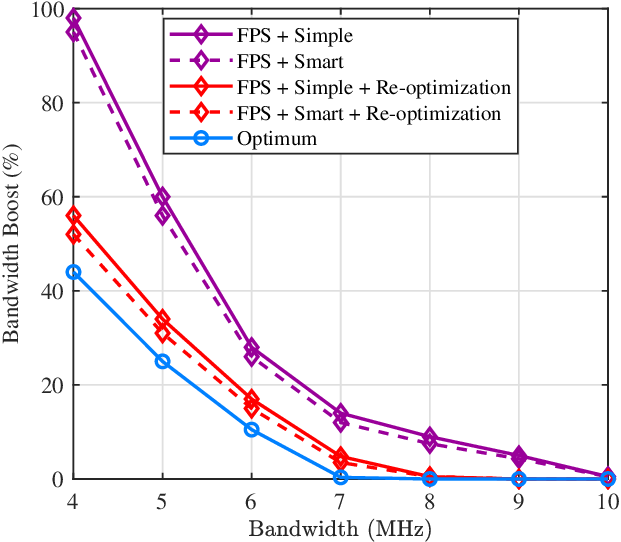}}
  \caption{Simulation Set I (FPS, P2): required bandwidth boost under varying demand $C_m$ and bandwidth $B$.}
  \label{fig:set1_fps_p2}
\end{figure}

\begin{figure}[H]
  \centering
  \subfloat[P3: energy-boost vs. $C_m$\label{set-one-P3-demand}]{
    \includegraphics[width=0.47\textwidth]{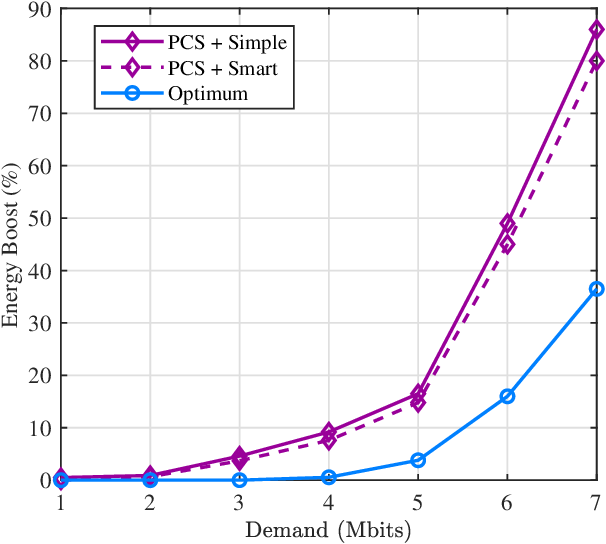}}\hfill
  \subfloat[P3: energy-boost vs. $E_m$\label{set-one-P3-energy}]{
    \includegraphics[width=0.47\textwidth]{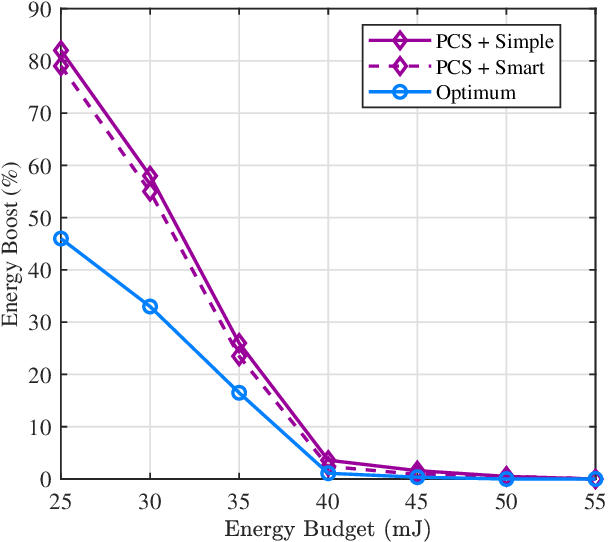}}
  \caption{Simulation Set I (PCS): required energy boost for P3 under varying demand $C_m$ and energy budget $E_m$.}
  \label{fig:set1_pcs}
\end{figure}

For each simulation run, the channel gains are averaged over 100 independent channel realizations, each generated from a distinct set of initial user locations, moving directions, and small-scale fading, and spanning for $T$ time slots. Based on the average channel gains, we apply the FPS Algorithm to P1 and P2 and the PCS algorithm to P3, and record the minimal bandwidth boost (FPS) or energy boost (PCS) required by the feasibility-boosting step to satisfy all constraints within the current $T$ time slots. The reported results represent an average over 100 independent simulation runs.
Since dependent rounding is inherently random, our results are obtained after five rounds of the rounding algorithms. Additional rounds further reduce required bandwidth or energy boost.
For benchmarking, we also obtained the Optimum solution by directly solving the original problems using Bonmin, where the bandwidth (for P1 and P2) and energy budget (for P3) is increased until the problem becomes feasible. Figures~\ref{set-one-P1-demand}-\ref{set-one-P3-energy} show the bandwidth or energy boost.

Figures~\ref{set-one-P1-demand} and~\ref{set-one-P1-bandwidth}, respectively, show the bandwidth boost when applying the FPS algorithm to P1, as the per-user demand $C_m$ and the original bandwidth $B$ change. 
Figure~\ref{set-one-P1-demand} shows that all methods require more bandwidth boost as the demand increases. 
Figure~\ref{set-one-P1-bandwidth} shows that as $B$ increases, the required bandwidth boost drops rapidly and then approaches  zero.
In both cases, Smart rounding slightly reduces the bandwidth boost compared with Simple rounding. Furthermore, for both rounding methods, the Re-optimization step significantly reduces the bandwidth boost and brings the requirement close to the optimum, though at the cost of additional decision latency.

Figures~\ref{set-one-P2-demand} and~\ref{set-one-P2-bandwidth}
respectively, show the bandwidth boost when applying the FPS algorithm to P2 as the per-user demand $C_m$ and the original bandwidth $B$ increase. 
Compared with the results in figures~\ref{set-one-P1-demand} and~\ref{set-one-P1-bandwidth},  P2 generally results in more bandwidth boosts, as the decisions are based on the channel gains over a shorter time interval and constraints must be satisfied within each block of $t_s$ time slots rather than across the entire $T$ time slots.

Figures~\ref{set-one-P3-demand} and~\ref{set-one-P3-energy}, respectively, show the required energy boosts  when applying the PCS algorithm to P3, as per-user demand $C_m$ and the energy budget $E_m$ change. 
As the demand increases, the required energy increases slowly at first and then more rapidly.  
As $E_m$ increases, the required energy boost decreases rapidly at the beginning and then gradually approaches zero once $E_m$ is sufficiently large.
Compared with Simple rounding, Smart rounding achieves slight reduction in  energy requirement. The gap between the PCS solutions and the optimum increases as $C_m$ increases or $E_m$ decreases. 

\subsection{Simulation Set II: scheduling performance}

In this set of simulations, we evaluate the performance of the resource allocation and scheduling decisions obtained from Set I. At the beginning of each $T$-interval, $100$ independent channel realizations are generated based on the current user locations and moving directions, and the average channel gains at each time slot are used to solve problems P1-P3. The runtime for solving the problem should be less than $T$ (for P1 and P3) or $t_s$ (for P2) time slots, and the corresponding  resource  allocation and scheduling decisions are applied to data transmissions in the subsequent $T$ (for P1 and P3) or $t_s$ (for P2) slots.   
For FPS (P1 and P2), the decisions include the boosted bandwidth $B^*$, the bandwidth allocation $w^*$; and for PCS (P3), the decisions include $T_f$, $x_r$, and the boosted per-user energy budgets. 
Since the resource allocation and scheduling decisions are not based on the exact channel gains, the demand constraints \cref{Eq:1C2} in \ref{Eq:1maxy} and \cref{Eq:2C2} in \ref{Eq:2maxy}, and the energy-budget constraint \cref{Eq:3C2} in \ref{Eq:3min0}, may be violated. For each of these constraints, the percentage of violation, defined as the difference between the two sides of the expression divided by the right-hand side, is recorded and then averaged over $100$ independent simulation runs.
The remaining constraints are not subject to violation, as their satisfaction is guaranteed by the solutions and unaffected by channel gain variations.

\begin{figure}[!t]
  \centering
  \subfloat[P1: violation vs. bandwidth\label{set-two-P1-bandwidth}]{
    \includegraphics[width=0.47\columnwidth]{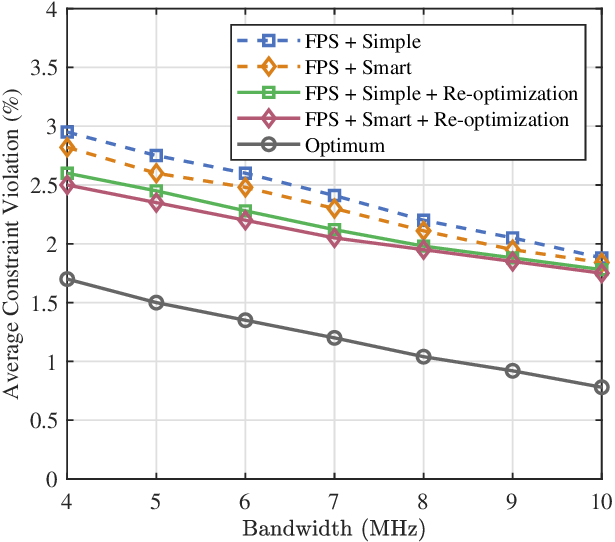}}\hfill
  \subfloat[P2: violation vs. bandwidth\label{set-two-P2-bandwidth}]{
    \includegraphics[width=0.47\columnwidth]{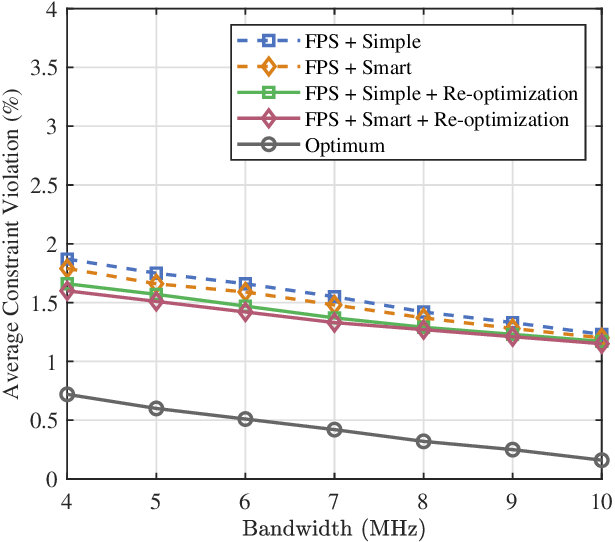}}\\[-0.25ex]
  \subfloat[P1: violation vs. demand\label{set-two-P1-demand}]{
    \includegraphics[width=0.47\columnwidth]{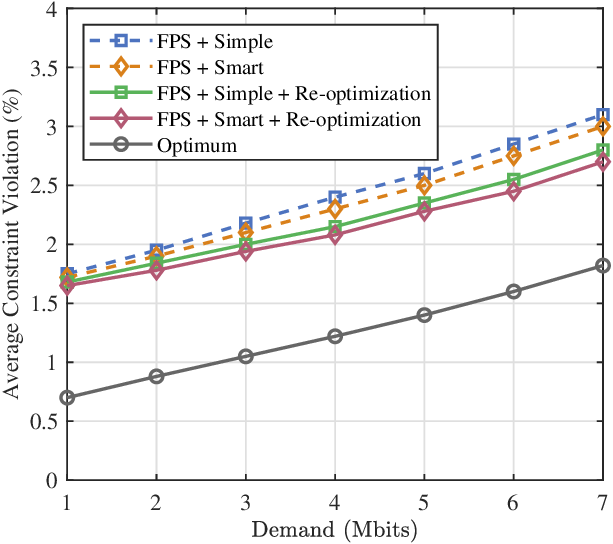}}\hfill
  \subfloat[P2: violation vs. demand\label{set-two-P2-demand}]{
    \includegraphics[width=0.47\columnwidth]{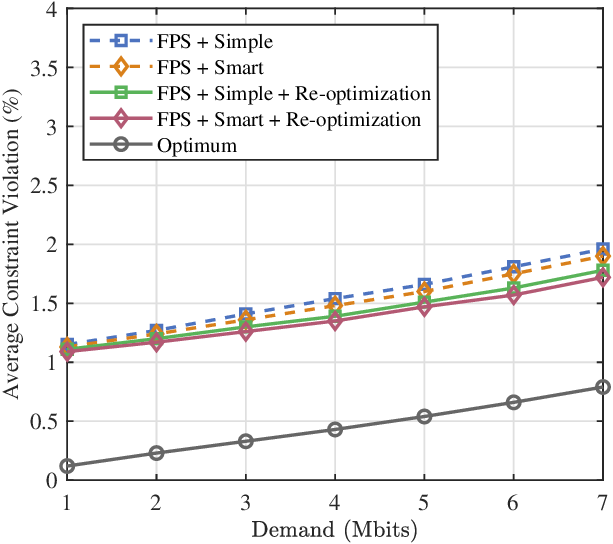}}
  \caption{Simulation Set II (FPS): average constraint violation for P1 and P2 under varying channel bandwidth $B$ and demand $C_m$.}
  \label{fig:set2_fps}
\end{figure}

Figures \ref{set-two-P1-bandwidth} and~\ref{set-two-P1-demand}, respectively, show the average constraint violation under the FPS algorithm for P1 as the bandwidth $B$ and demand $C_m$ change. 
The results indicate that the constraint violations remain relatively low across all solution methods. Smart rounding achieves lower constraint violations than Simple rounding, and incorporating Re-optimization with either Simple or Smart rounding further reduces the violations. The violation rate achieved by  Smart rounding with re-optimization is within approximately 1\% higher than the Optimum, whereas Simple rounding without Re-optimization is about 2\% above the optimum. Since the Optimum solution is obtained using the exact channel gains over the entire $T$ time slots (for P1 and P3) or $t_s$ time slots (for P2), the proposed solutions demonstrate very strong performance in comparison.

Figures \ref{set-two-P2-bandwidth} and~\ref{set-two-P2-demand}, respectively, show the average constraint violation under the FPS algorithm for P2 as the bandwidth $B$ and demand $C_m$ change. 
Similar to the results for P1 in figures \ref{set-two-P1-bandwidth} and~\ref{set-two-P1-demand}, the violations remain low and close to the Optimum across all proposed solutions. A comparison with figures \ref{set-two-P1-bandwidth} and~\ref{set-two-P1-demand} further indicates that  the violations using P2 are smaller than using P1, and this is due to higher bandwidth boosts using P2.

\begin{figure*}[!t]
  \centering
  \subfloat[P3: violation vs. energy budget $E_m$\label{set-two-P3-energy}]{
    \includegraphics[width=0.48\textwidth]{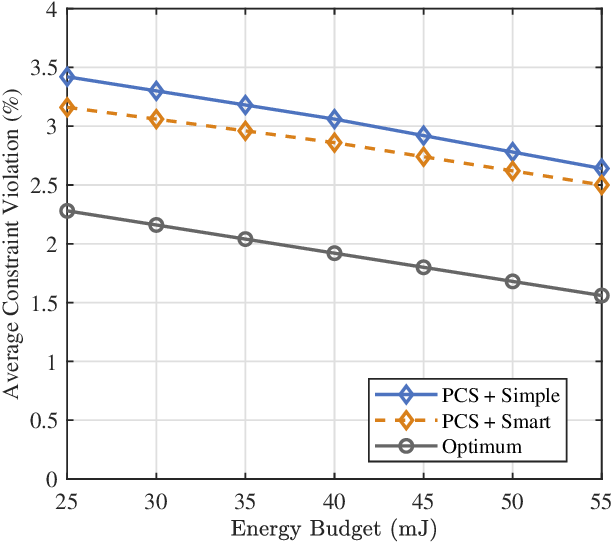}}\hfill
  \subfloat[P3: violation vs. demand $C_m$\label{set-two-P3-demand}]{
    \includegraphics[width=0.48\textwidth]{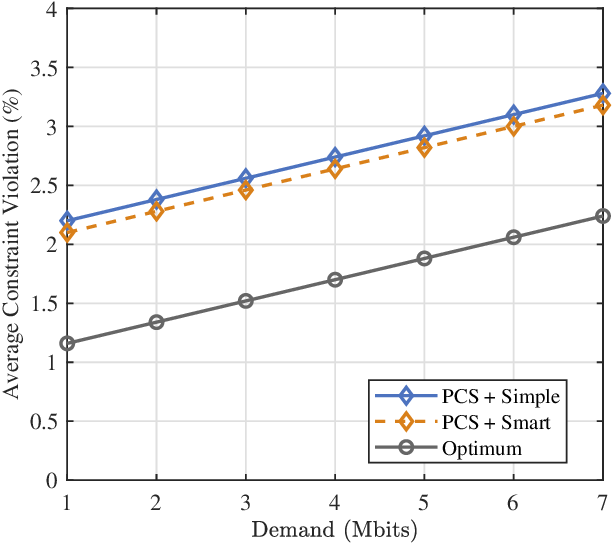}}
  \caption{Simulation Set II (PCS): average constraint violation for P3 under varying energy budget $E_m$ and demand $C_m$.}
  \label{fig:set2_pcs}
\end{figure*}

Figures \ref{set-two-P3-energy} and  \ref{set-two-P3-demand}, respectively, show the average constraint violation using the PCS algorithm for P3 as the energy budget $E_m$ and demand $C_m$ change. The constraint violation is consistently low and close to the optimum.

Overall, we can conclude that i) Smart rounding achieves lower resource boosts (bandwidth or energy) and reduced constraint violations compared to Simple rounding, while maintaining similar complexity; ii) Re-optimization significantly reduces the bandwidth boosts in both P1 and P2; and iii) all proposed solutions achieve constraint violation close to the optimum. 
In practice, the user may continue transmitting the remaining bits at the end of the $T_f$ interval. Since the constraint violations are small, data transmissions can be completed with a few additional time slots (typically one or two). Therefore, as long as $T>T_f$, all user data can be successfully delivered within $T$ time slots.

In all of the presented results, the performance of the proposed algorithms have been compared with an optimum result obtained by directly solving the original problem formulations. It can be seen that the performance gap compared to this optimum is small, and the violation percentages are small values, indicating that the proposed method achieves near-optimal schedules.

\subsection{Simulation Set III: algorithm runtime}
\label{subsec:latency}

\cref{tab:runtime} reports the mean runtime of the different solution methods.
 For P1 and P2, ``Base'' denotes the runtime of FPS with either Simple or Smart rounding (without Re-optimization), as the two rounding methods result in nearly identical runtimes. ``+ Re-Opt'' is the runtime of the Base method plus the additional runtime required for the Re-optimization stage. 
For P3, ``Base'' denotes the runtime of PCS with either Simple or Smart rounding. 
As shown in the table, enabling Re-optimization increases the overall runtime. 
The relative runtimes are consistent with the problem sizes and solver call patterns. P1 is the slowest because it conducts a binary search on $T$ and therefore invokes the federated solver multiple times; P2 is faster than P1 because it optimizes over a fixed block of $t_s$ slots with fewer decision variables; and P3 is faster than P1 since its fractional relaxation reduces to a linear program that can be solved efficiently. We note that the iterative federated optimization is done via the backbone network that connects the DT edge servers. Therefore, this communication is a small part of the measured runtime in \cref{tab:runtime}, i.e., the total scheduling latency is dominated by the computation of the fractional optimal solutions.
Moreover, since the user group size is kept small as mentioned earlier, the problem size allows the algorithms to run  within the current scheduled period. Scaling to larger user populations is achieved by increasing the number of user groups.

For timing guarantees, recall that P1 and P3 operate over a period of $T$ slots of duration $\tau$, and the schedule computed for the next period must be available before the current period ends so it can be disseminated to users and applied at the beginning of the next period (one-period look-ahead operation). Therefore, the execution time of the algorithm should not exceed $T \tau = 500$ ms. 
For P2, the runtime reported in \cref{tab:runtime} is the total runtime for an entire $T$-interval, comprised of a sequence of $t_s$-blocks. Since each $t_s$-block is used to produce the schedule for the upcoming $t_s$ slots, the scheduler must be done in fewer than $t_s$ time slots. Indeed, the runtime of P2 is between 10ms and 14ms, which is well below the limit of $t_s \tau = 50$ ms.

All measurements were obtained on a desktop with an Intel\textsuperscript{\textregistered} Core i7 12700 processor with 12 cores and 20 threads at 2.10\,GHz and 64\,GB of memory on a 64 bit operating system. In practical deployments, each user’s DT may run on an edge server with newer processors and more cores. Since client-side work in each federated round parallelizes across the DTs, such multi-core edge hardware can achieve shorter wall-clock times than our desktop baseline, so actual execution times can be considerably better than those shown in the table.

\begin{table}[t]
  \centering
  \caption{Runtime of the proposed solver.}
  \label{tab:runtime}
  \vspace{2pt}
  \footnotesize
  \setlength{\tabcolsep}{6pt} 
  \begin{tabular}{@{}ccccc@{}}
    \cmidrule(l{-6pt}r{-6pt}){1-5}
    \multicolumn{2}{c}{\textbf{Problem P1}} &
    \multicolumn{2}{c}{\textbf{Problem P2}} &
    \textbf{P3} \\
    \cmidrule(l{-6pt}r{-6pt}){1-2}
    \cmidrule(l{-6pt}r{-6pt}){3-4}
    \cmidrule(l{-6pt}r{-6pt}){5-5}
    Base & +\,Re-Opt & Base & +\,Re-Opt & Base \\
    \cmidrule(l{-6pt}r{-6pt}){1-5}
    $125$ ms & $180$ ms & $50$ ms & $70$ ms & $60$ ms \\
    \cmidrule(l{-6pt}r{-6pt}){1-5}
  \end{tabular}
  \vspace{-2mm}
\end{table}

\section{Concluding remarks}
DTs may protect information that is considered private to their associated physical systems. For a mobile device, this may include things such as its mobility profile, recent location(s), and experienced channel conditions. Online schedulers typically use this type of information to perform tasks such as shared bandwidth and channel time slot assignments. In this paper, we considered three transmission scheduling problems with energy constraints, where such information is needed, and yet must remain private: minimizing total transmission time when (i) fixed-power or (ii) fixed-rate time slotting with power control is used, and (iii) maximizing the amount of data uploaded in a fixed time period. Using a real-time federated optimization framework, we showed how the scheduler can iteratively interact only with the DTs to produce global fractional solutions to these problems, without the latter revealing their private information. Then dependent rounding was used to round the fractional solution into a channel transmission schedule for the physical systems. 
Experiments on wireless traces showed that consistent makespan reductions with near-zero bandwidth/energy violations are possible with millisecond-order end-to-end runtime for typical edge server hardware. To the best of our knowledge, this is the first framework that enables channel sharing across DTs using operations that do not expose private data.

\bibliographystyle{plain}   
\bibliography{main}

\begin{thebibliography}{10}

\bibitem{bonmin}
Bonmin (basic open-source mixed integer programming).
\newblock \url{http://projects.coin-or.org/Bonmin}.

\bibitem{Aybat2016NIPS}
Necdet~S. Aybat and Erfan~Y. Hamedani.
\newblock A primal-dual method for conic constrained distributed optimization
  problems.
\newblock In {\em Advances in Neural Information Processing Systems (NeurIPS)},
  volume~29, 2016.

\bibitem{Aybat2019SIOPT}
Necdet~S. Aybat and Erfan~Y. Hamedani.
\newblock A distributed admm-like method for resource sharing over time-varying
  networks.
\newblock {\em SIAM Journal on Optimization}, 29(4):3036--3068, 2019.

\bibitem{bonami2008algorithmic}
P.~Bonami, L.~T. Biegler, A.~R. Conn, G.~Cornuéjols, I.~E. Grossmann, C.~D.
  Laird, J.~Lee, A.~Lodi, F.~Margot, N.~Sawaya, and A.~Wächter.
\newblock An algorithmic framework for convex mixed integer nonlinear programs.
\newblock {\em Discrete Optimization}, 5(2):186--204, 2008.

\bibitem{Chen2024DT_Microservice}
H.~Chen, X.~Liu, Y.~Wang, Z.~Li, and S.~Wang.
\newblock Digital twin–assisted microservice offloading and bandwidth
  allocation via reinforcement learning.
\newblock arXiv preprint arXiv:2403.08687, Mar 2024.

\bibitem{Chen2024}
H.~Chen, T.~D. Todd, D.~Zhao, and G.~Karakostas.
\newblock Digital twin model selection for feature accuracy.
\newblock {\em IEEE Internet of Things Journal}, 11(7):11415--11426, Apr 2024.

\bibitem{Chu2021FedFair}
Lingyang Chu, Lanjun Wang, Yanjie Dong, Jian Pei, Zirui Zhou, and Yong Zhang.
\newblock Fedfair: Training fair models in cross-silo federated learning.
\newblock {\em arXiv preprint arXiv:2109.05662}, 2021.

\bibitem{Cunha2025RSSIPrivacy}
M.~Cunha et~al.
\newblock Compromising location privacy through wi-{Fi} {RSSI} tracking.
\newblock {\em Scientific Reports}, 2025.

\bibitem{Dieuleveut2021FedEM}
Aymeric Dieuleveut, Gersende Fort, Eric Moulines, and Genevi\`eve Robin.
\newblock Federated expectation maximization with heterogeneity mitigation and
  variance reduction.
\newblock In {\em NeurIPS}, 2021.

\bibitem{Carmo2024Survey_DTOffload}
P.~R.~X. do~Carmo, D.~F. Bezerra, A.~O. Filho, and E.~Freitas.
\newblock Living on the edge: A survey of digital twin–assisted task
  offloading in safety-critical environments.
\newblock {\em J. Netw. Comput. Appl.}, 232:104024, 2024.

\bibitem{Du2021SDM}
Wei Du, Depeng Xu, Xintao Wu, and Hanghang Tong.
\newblock Fairness-aware agnostic federated learning.
\newblock In {\em SIAM International Conference on Data Mining (SDM)}, pages
  181--189, 2021.

\bibitem{Galvez2021NeurIPSW}
Borja~R. G{\'a}lvez, Filip Granqvist, Rogier van Dalen, and Matt Seigel.
\newblock Enforcing fairness in private federated learning via the modified
  method of differential multipliers.
\newblock In {\em NeurIPS Workshop on Privacy in Machine Learning}, 2021.

\bibitem{Gandhi2006}
R.~Gandhi, S.~Khuller, S.~Parthasarathy, and A.~Srinivasan.
\newblock Dependent rounding and its applications to approximation algorithms.
\newblock {\em Journal of the ACM (JACM)}, 53(3):324--360, 2006.

\bibitem{Hao2023JSAC_DT_URLLC}
Y.~Hao, J.~Wang, D.~Huo, N.~Guizani, L.~Hu, and M.~Chen.
\newblock Digital twin-assisted {URLLC}-enabled task offloading in mobile edge
  network via robust combinatorial optimization.
\newblock {\em IEEE J. Sel. Areas Commun.}, 41(10):3022--3036, Oct 2023.

\bibitem{He2024FederatedConstraints}
C.~He, L.~Peng, and J.~Sun.
\newblock Federated learning with convex global and local constraints.
\newblock {\em Transactions on Machine Learning Research}, 2024.
\newblock \url{https://openreview.net/forum?id=qItxVbWyfe}.

\bibitem{He2023DTFL_HCN}
Y.~He, M.~Yang, Z.~He, and M.~Guizani.
\newblock Resource allocation based on digital twin-enabled federated learning
  framework in heterogeneous cellular network.
\newblock {\em IEEE Trans. Veh. Technol.}, 72(1):1149--1158, Jan 2023.

\bibitem{HeydariVTC2025}
Mohammad Heydari, Terence~D. Todd, Dongmei Zhao, and George Karakostas.
\newblock Scheduling and resource allocation for federated learning in
  vehicular networks.
\newblock In {\em 2025 IEEE 102nd Vehicular Technology Conference
  (VTC2025-Fall)}, pages 1--6, 2025.

\bibitem{Huang2024UDTResourceMgmt}
Xinyu Huang, Wen Wu, Shisheng Hu, Mushu Li, Conghao Zhou, and Xuemin Shen.
\newblock Digital twin based user-centric resource management for multicast
  short video streaming.
\newblock {\em IEEE Journal of Selected Topics in Signal Processing},
  18(1):50--65, 2024.

\bibitem{jakes1994microwave}
William~C Jakes and Donald~C Cox.
\newblock {\em Microwave mobile communications}.
\newblock Wiley-IEEE press, 1994.

\bibitem{Kairouz2021FL}
Peter Kairouz, H.~Brendan McMahan, Brendan Avent, Aur{\'e}lien Bellet, Mehdi
  Bennis, Arjun~Nitin Bhagoji, Kallista Bonawitz, Zachary Charles, Graham
  Cormode, Rachel Cummings, Rafael G.~L. D'Oliveira, Hubert Eichner, Salim
  El~Rouayheb, David Evans, Josh Gardner, Zachary Garrett, Adri{\`a}
  Gasc{\'o}n, Badih Ghazi, Phillip~B. Gibbons, Marco Gruteser, Zaid Harchaoui,
  Chaoyang He, Lie He, Zhouyuan Huo, Ben Hutchinson, Justin Hsu, Martin Jaggi,
  Tara Javidi, Gauri Joshi, Mikhail Khodak, Jakub Kone{\v{c}}n{\'y}, Aleksandra
  Korolova, Farinaz Koushanfar, Sanmi Koyejo, Tancr{\`e}de Lepoint, Yang Liu,
  Prateek Mittal, Mehryar Mohri, Richard Nock, Ayfer {\"O}zg{\"u}r, Rasmus
  Pagh, Hang Qi, Daniel Ramage, Ramesh Raskar, Mariana Raykova, Dawn Song,
  Weikang Song, Sebastian~U. Stich, Ziteng Sun, Ananda~Theertha Suresh, Florian
  Tram{\`e}r, Praneeth Vepakomma, Jianyu Wang, Li~Xiong, Zheng Xu, Qiang Yang,
  Felix~X. Yu, Han Yu, and Sen Zhao.
\newblock Advances and open problems in federated learning.
\newblock {\em Foundations and Trends in Machine Learning}, 14(1--2):1--210,
  2021.

\bibitem{Kim2022}
J.~Kim, S.~Hosseinalipour, A.~C. Marcum, T.~Kim, D.~J. Love, and C.~G. Brinton.
\newblock Learning-based adaptive irs control with limited feedback codebooks.
\newblock {\em IEEE Trans. Wireless Commun.}, 21(11):9566--9581, Nov 2022.

\bibitem{Li2022TVT_DT_UAV_Offload}
B.~Li, Y.~Liu, L.~Tan, H.~Pan, and Y.~Zhang.
\newblock Digital twin assisted task offloading for aerial edge computing and
  networks.
\newblock {\em IEEE Trans. Veh. Technol.}, 71(10):10863--10877, Oct 2022.

\bibitem{LiMitraTSP2024FPI}
Jianxiu Li and Urbashi Mitra.
\newblock Channel state information-free location-privacy enhancement: Fake
  path injection.
\newblock {\em IEEE Transactions on Signal Processing}, 72:3745--3760, 2024.

\bibitem{Li2024DTResourceAlloc}
Mushu Li, Jie Gao, Conghao Zhou, Lian Zhao, and Xuemin Shen.
\newblock Digital-twin-empowered resource allocation for on-demand
  collaborative sensing.
\newblock {\em IEEE Internet of Things Journal}, 11(23):37942--37958, 2024.

\bibitem{Lin2016Automatica}
Peng Lin, Wei Ren, and Yongduan Song.
\newblock Distributed multi-agent optimization subject to nonidentical
  constraints and communication delays.
\newblock {\em Automatica}, 65:120--131, 2016.

\bibitem{Liu2022AGV_IIoT_Offloading}
Peng Liu, Zhe Liu, Ji~Wang, Zifu Wu, Peng Li, and Huijuan Lu.
\newblock Reinforcement learning empowered multi-agv offloading scheduling in
  edge-cloud iiot.
\newblock {\em Journal of Cloud Computing}, 11(1):78, 2022.

\bibitem{Lu2021LowLatencyDTFL}
Y.~Lu, X.~Huang, K.~Zhang, S.~Maharjan, and Y.~Zhang.
\newblock Low-latency federated learning and blockchain for edge association in
  digital twin empowered 6g networks.
\newblock {\em IEEE Trans. Ind. Inf.}, 17(7):5098--5107, Jul 2021.

\bibitem{materwala2022energy}
Huned Materwala, Leila Ismail, Raed~M Shubair, and Rajkumar Buyya.
\newblock Energy-sla-aware genetic algorithm for edge--cloud integrated
  computation offloading in vehicular networks.
\newblock {\em Future Gener. Comput. Syst.}, 135:205--222, 2022.

\bibitem{mcmahan2017communication}
B.~McMahan, E.~Moore, D.~Ramage, S.~Hampson, and B.~A. Arcas.
\newblock Communication-efficient learning of deep networks from decentralized
  data.
\newblock In {\em Proc. 20th Int. Conf. Artif. Intell. Statist. (PMLR)},
  volume~54, pages 1273--1282, Apr 2017.

\bibitem{Nedic2010TAC}
A.~Nedi\'c, A.~Ozdaglar, and P.~A. Parrilo.
\newblock Constrained consensus and optimization in multi-agent networks.
\newblock {\em IEEE Trans. Autom. Control}, 55(4):922--938, 2010.

\bibitem{Shen2021ICLR}
Zebang Shen, Juan Cervino, Hamed Hassani, and Alejandro Ribeiro.
\newblock An agnostic approach to federated learning with class imbalance.
\newblock In {\em International Conference on Learning Representations (ICLR)},
  2021.

\bibitem{Sklar2001}
B.~Sklar.
\newblock {\em Digital Communications: Fundamentals and Applications}.
\newblock Prentice Hall, 2nd edition, 2001.

\bibitem{Son2022PrivacyPreservingDT}
Seunghwan Son, Deokkyu Kwon, Joonyoung Lee, Sungjin Yu, Nam-Su Jho, and Youngho
  Park.
\newblock On the design of a privacy-preserving communication scheme for
  cloud-based digital twin environments using blockchain.
\newblock {\em IEEE Access}, 10:75365--75375, July 2022.

\bibitem{Sun2021DTIIoT}
W.~Sun, S.~Lei, L.~Wang, Z.~Liu, and Y.~Zhang.
\newblock Adaptive federated learning and digital twin for industrial internet
  of things.
\newblock {\em IEEE Trans. Ind. Inf.}, 17(8):5605--5614, Aug 2021.

\bibitem{Vaezi2022NetworkingPerspective}
Mehrad Vaezi, Kiana Noroozi, Terence~D. Todd, Dongmei Zhao, George Karakostas,
  Huaqing Wu, and Xuemin Shen.
\newblock Digital twins from a networking perspective.
\newblock {\em IEEE Internet of Things Journal}, 9(23):23525--23544, December
  2022.

\bibitem{Wang2017TAC}
Peng Wang, Peng Lin, Wei Ren, and Yongduan Song.
\newblock Distributed subgradient-based multiagent optimization with more
  general step sizes.
\newblock {\em IEEE Transactions on Automatic Control}, 63(7):2295--2302, 2018.

\bibitem{Wu2025DTFL_NOMA}
B.~Wu, F.~Fang, and X.~Wang.
\newblock Stackelberg game based performance optimization in digital twin
  assisted federated learning over noma networks.
\newblock arXiv preprint arXiv:2501.01584, Jan 2025.

\bibitem{Yang2019ARC}
Tao Yang, Xinlei Yi, Junfeng Wu, Ye~Yuan, Di~Wu, Ziyang Meng, Yiguang Hong,
  Hong Wang, Zongli Lin, and Karl~H. Johansson.
\newblock A survey of distributed optimization.
\newblock {\em Annual Reviews in Control}, 47:278--305, 2019.

\bibitem{Yang2023DTIndustrial}
W.~Yang, W.~Xiang, Y.~Yang, and P.~Cheng.
\newblock Optimizing federated learning with deep reinforcement learning for
  digital twin empowered industrial iot.
\newblock {\em IEEE Trans. Ind. Inf.}, 19(2):1884--1893, Feb 2023.

\bibitem{Yuan2011TSMCB}
Deming Yuan, Shengyuan Xu, and Huanyu Zhao.
\newblock Distributed primal--dual subgradient method for multiagent
  optimization via consensus algorithms.
\newblock {\em IEEE Transactions on Systems, Man, and Cybernetics, Part B
  (Cybernetics)}, 41(6):1715--1724, 2011.

\bibitem{Zhang2024DTFL_MEN}
R.~Zhang, Z.~Xie, D.~Yu, W.~Liang, and X.~Cheng.
\newblock Digital twin-assisted federated learning service provisioning over
  mobile edge networks.
\newblock {\em IEEE Trans. Comput.}, 73(2):586--598, Feb 2024.

\bibitem{Zhang2023JSAC_DT_Offloading}
Y.~Zhang, Z.~Xie, D.~Yu, W.~Liang, and X.~Cheng.
\newblock Digital twin–driven intelligent task offloading for collaborative
  mobile edge computing.
\newblock {\em IEEE J. Sel. Areas Commun.}, 41(10), Oct 2023.

\bibitem{Zhu2019DLG}
L.~Zhu, Z.~Liu, and S.~Han.
\newblock Deep leakage from gradients.
\newblock In {\em Advances in Neural Information Processing Systems 32
  (NeurIPS)}, pages 14774--14784, Dec 2019.

\bibitem{Zhu2011TAC}
Minghui Zhu and Sonia Mart{\'\i}nez.
\newblock On distributed convex optimization under inequality and equality
  constraints.
\newblock {\em IEEE Transactions on Automatic Control}, 57(1):151--164, 2012.

\end{thebibliography}

\appendix
\section*{Appendix}

In \cref{apx:ad-nr}, we include some extra numerical results.

\section{Extra Numerical Results}\label{apx:ad-nr}
\subsection{Robustness to Mobility: Sensitivity to User Speed}\label{apx:dataset-desc}

\begin{figure*}[!htbp]
  \centering
  \subfloat[FPS (P1 \& P2): average constraint violation vs. user speed\label{set-two-P1P2-speed}]{
    \includegraphics[width=0.48\textwidth]{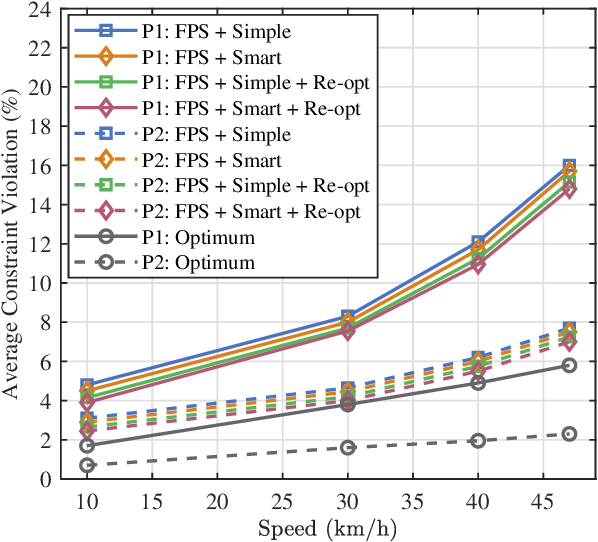}}\hfill
  \subfloat[PCS (P3): average constraint violation vs. user speed\label{set-two-P3-speed}]{
    \includegraphics[width=0.48\textwidth]{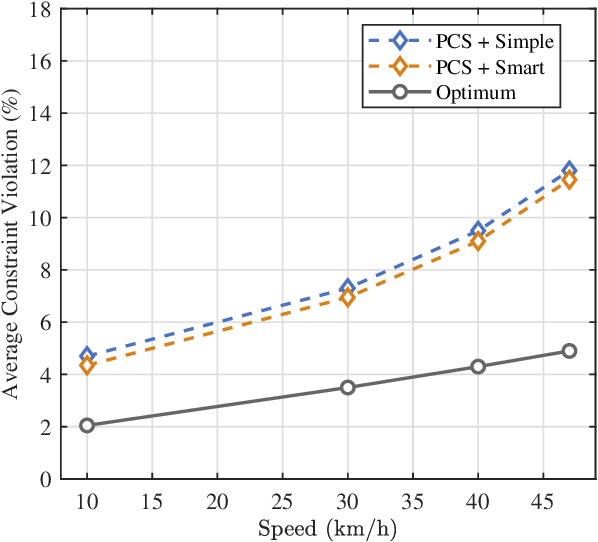}}
  \caption{Speed sensitivity on empirical channel traces. Panels show FPS (left) and PCS (right).}
  \label{fig:app_speed_twopanel}
\end{figure*}

Figures~\ref{set-two-P1P2-speed} and~\ref{set-two-P3-speed} show the constraint violation as the user moving speed changes. As speed increases, channel correlation decreases, and the resource allocation and scheduling decisions based on past channel gains become less accurate when applied to the current time period. As a result, constraint violations increase with speed for all the solution methods. 
Figure~\ref{set-two-P1P2-speed} further shows that P2 consistently yields lower constraint violations than P1 for all the solution methods, and the gap between the two increases as moving speed increases. Overall, the solutions to P2
are much less sensitive to the moving speed than those of P1, since P2 is solved over shorter time intervals.
The difference between each proposed solution and the optimum also increases with the moving speed. The main reason is that we adopted the same time slot duration for all the experiments. In practice, as the moving speed increases, channel coherence time decreases, the resource and scheduling decisions should be updated more frequently to reduce constraint violation.

\subsection{Runtime Sensitivity to Accuracy Tolerances \texorpdfstring{\((\epsilon_1,\epsilon_2)\)}{}}
\label{sec:runtime-tolerances}

We quantify how solver runtime varies with the accuracy tolerances \(\boldsymbol{\epsilon}_1\) and \(\boldsymbol{\epsilon}_2\). For each problem (P1--P3), we sweep \((\boldsymbol{\epsilon}_1,\boldsymbol{\epsilon}_2)\) over a fixed grid while holding all other settings constant, and report mean runtime over repeated trials on the same hardware. As expected, tighter tolerances increase runtime, illustrating the accuracy--latency trade-off in our configurations. Results are summarized in \cref{tab:runtime-appendix}.

\begin{table}[H]
  \centering
  \caption{Mean runtime of the proposed solver under varying accuracy tolerances $(\epsilon_{1},\epsilon_{2})$. 
           “+\,Re-Opt” denotes execution with the re-optimization stage enabled.}
  \label{tab:runtime-appendix}
  \vspace{2pt}
  \setlength{\tabcolsep}{1.5pt}
  \begin{tabular}{@{}cccccccc@{}}
    \toprule
    \multicolumn{2}{c}{Tolerance parameters} &
    \multicolumn{2}{c}{Problem P1} &
    \multicolumn{2}{c}{Problem P2} &
    \multicolumn{1}{c}{\quad P3 \quad} \\[-0.2ex]
    \cmidrule(lr){1-2} \cmidrule(lr){3-4} \cmidrule(lr){5-6} \cmidrule(lr){7-7}
    $\epsilon_{1}$ & $\epsilon_{2}$ & Base & +\,Re-Opt & Base & +\,Re-Opt & Base \\ \midrule
    $0.5$ & $0.05$ & $125$ ms & $180$ ms & $50$ ms & $70$ ms & $60$ ms \\
    $10^{-2}$        & $10^{-2}$        & $600$ ms & $850$ ms & $175$ ms & $220$ ms & $250$ ms \\
    $10^{-3}$        & $10^{-3}$        & $4000$ ms & $5800$ ms & $1250$ ms & $1600$ ms & $1900$ ms \\
    $10^{-4}$        & $10^{-4}$        & $30000$ ms & $44000$ ms & $9200$ ms & $13000$ ms & $16000$ ms \\ \bottomrule
  \end{tabular}
  \vspace{-2mm} 
\end{table}

\end{document}